\documentclass[12pt]{elsarticle}                                          
\usepackage{graphicx}                                                         
\usepackage{a41}                                                                
\usepackage{xcolor}                     
\usepackage[rflt]{floatflt}
\usepackage{float}
\usepackage{lscape}
\usepackage{breakurl} 
\usepackage{times}
\usepackage{array}
\usepackage[english]{babel}
\usepackage[T1]{fontenc}
\usepackage{ae}
\usepackage{url}
\usepackage{amsmath, amsthm, amssymb}
\usepackage{slashed}

\usepackage{rotating}
\usepackage{graphicx}
\usepackage{comment}
\newcounter{mmacnt}
\def\restartmma{\setcounter{mmacnt}{0}}
\restartmma \catcode`|=\active
\def|#1|{\mathrm{#1}}
\catcode`|=12
\newenvironment{mma}{
\par\smallskip
\catcode`|=\active
\parskip=0pt\parindent=0pt 
\small
\def\In##1\\{%
\def\linebreak{\hfill\break\null\qquad}%
\refstepcounter{mmacnt}
\hangindent=2.5em\hangafter=0
\leavevmode
\llap{\tiny\sffamily In[\arabic{mmacnt}]:=\kern.5em}%
\mathversion{bold}\footnotesize$
\displaystyle##1$\normalsize
\mathversion{normal}\par
 }%
\def\Print##1\\{%
\def\linebreak{\hfill\break}%
\hangindent=2.5em\hangafter=0
\leavevmode ##1\par}%
\def\Out##1\\{%
\def\linebreak{$\hfill\break\null\hfill$}%
\kern\abovedisplayskip\par
\hangindent=2.5em\hangafter=0
\leavevmode
\llap{\tiny\sffamily Out[\arabic{mmacnt}]=\kern.5em}
\footnotesize$\displaystyle##1$
\normalsize\hfill\null\par
\kern\belowdisplayskip
}%
\def\Warning##1##2\\{%
\def\linebreak{\hfill\break}%
\hangindent=2.5em\hangafter=0
\leavevmode
{\scriptsize##1 : ##2}\par}%
}{%
\par\smallskip
}

\usepackage{color}
\newenvironment{fshaded}{%
\MakeFramed {\FrameRestore}
}%
{\endMakeFramed}

\makeatletter
\def\ps@pprintTitle{%
\let\@oddhead\@empty
\let\@evenhead\@empty
\def\@oddfoot{\reset@font\hfil\thepage\hfil}
\let\@evenfoot\@oddfoot
}
\makeatother
\usepackage{tikz}
\usetikzlibrary{matrix}
\allowdisplaybreaks[4]

\begin{document}
\begin{frontmatter}
\title{\Large
\textbf{
One-loop expressions for 
$H^{\pm} \rightarrow W^{\pm} Z$
and their implications at 
muon--TeV colliders
}}
\author[1,2]{Dzung Tri Tran}
\author[3]{Quang Hoang-Minh Pham}
\author[3]{Khoa Ngo-Thanh Ho}
\author[1,2]{Khiem Hong Phan}
\ead{phanhongkhiem@duytan.edu.vn}
\address[1]{\it Institute of Fundamental
and Applied Sciences, Duy Tan University,
Ho Chi Minh City $70000$, Vietnam}
\address[2]{Faculty of Natural Sciences,
Duy Tan University, Da Nang City $50000$,
Vietnam}
\address[3]
{\it VNUHCM-University of Science,
$227$ Nguyen Van Cu, District $5$,
Ho Chi Minh City $700000$, Vietnam}
\pagestyle{myheadings}
\markright{}
\begin{abstract} 
One-loop contributions for decay process
$H^{\pm} \rightarrow W^{\pm}Z$
within the Two-Higgs-Doublet Model is computed in
the general $\mathcal{R}_{\xi}$ gauge, and its
phenomenological applications at future muon--TeV
colliders are studied in this paper. The analytic
results are confirmed by several consistency tests,
for example, the $\xi$-independence, the
renormalization-scale stability
and the ultraviolet finiteness of the
one-loop amplitude. We first perform an
updated parameter scan of the Type-X THDM in the
phenomenological studies. The
production of charged Higgs boson pairs
at future muon--TeV colliders
is investigated through the two processes $\mu^+\mu^-
\rightarrow H^+H^- \rightarrow W^{\pm}W^{\mp}Zh$ and $\mu^+\mu^-
\rightarrow \gamma\gamma \rightarrow H^+H^- \rightarrow
W^{\pm}W^{\mp}Zh$. Both signal events and their significances
are evaluated with taking into account the corresponding Standard
Model backgrounds. We find that the signal significances
can exceed $5\sigma$ at several benchmark points in the viable
parameter space of the Type-X THDM.
\end{abstract}
\end{frontmatter}
\section{Introduction}
Searches for additional scalar Higgs
boson production in many extensions
of the Standard Model (SM) are one
of the main purposes of future colliders,
including the High-Luminosity LHC (HL-LHC)
and proposed lepton colliders such as
the International Linear Collider
(ILC) and muon--TeV colliders.
The discovery of the additional scalar Higgs
bosons would provide direct evidence for
new physics and would also offer
a enhanced insight into the dynamics of
electroweak symmetry breaking (EWSB).
In all possible production channels of exotic
scalar states, singly charged Higgs boson
production has recently received particular
attention at colliders. Experimentally, searches
for charged Higgs bosons in the light
mass region, produced in top-quark decays
have been detected at $\sqrt{s}=7$ and $8$~TeV
at the LHC as in Refs.~\cite{CMS:2012fgz,CMS:2015lsf,
ATLAS:2023bzb,ATLAS:2024oqu}.
Additional exploration of charged Higgs bosons
following decay channels
$H^{+} \to \tau\nu$~\cite{ATLAS:2012nhc}
and $H^+ \to c\bar{s}$~\cite{ATLAS:2013uxj},
have also been examined by the ATLAS
Collaboration. For heavier charged Higgs
states, both ATLAS and CMS have conducted
searches through decay channels such as
$H^{\pm} \to tb$~\cite{ATLAS:2015nkq,
ATLAS:2021upq,CMS:2020imj}, $H^{\pm} \to
W^{\pm}Z$~\cite{ATLAS:2015edr,
ATLAS:2018iui} at $\sqrt{s}=8$~TeV, and via
vector boson fusion production at
$\sqrt{s}=13$~TeV~\cite{CMS:2021wlt}.
Furthermore, searches for $H^\pm
\to cb/cs$ at $\sqrt{s}=8$~TeV have been reported in
Refs.~\cite{CMS:2018dzl,CMS:2020osd}, while the
$H^{\pm}\to HW^{\pm}$ decay mode has been
studied at $\sqrt{s}=13$~TeV~\cite{CMS:2022jqc,
ATLAS:2024rcu}. More recently, both ATLAS and
CMS have carried out the investigations
for charged Higgs bosons in association with
top quarks and in top-quark decays, in which both
production channels have been analyzed
with the subsequent decay $H^{\pm}\to \tau^{\pm}
\nu_{\tau}$~\cite{ATLAS:2018gfm,ATLAS:2024hya,
CMS:2019bfg}.

Theoretically, charged Higgs boson production
at the LHC has been calculated within many BSM
scenarios. In the THDM, $pp \to tH^- \to
tW^-b\bar{b}$ production has been computed,
including top-quark polarization effects as
discussed in~\cite{Arhrib:2017veb,Arhrib:2018bxc}.
Additionally, the decay $H^+ \to t\bar{b}$ has
been systematically examined in the
MSSM~\cite{Arhrib:2019ykh}.
The production of charged Higgs boson pairs
at the HL-LHC has been investigated
in~\cite{Arhrib:2019ywg}.
Taking Run~III data, charged Higgs bosons in
the light-mass region decaying into electroweak
vector bosons have been analyzed in the
report~\cite{Arhrib:2020tqk}.
Investigations of the productions for $pp \to H^{\pm}h/A$
and $pp \to H^{+}H^{-}$ with $H^{\pm}\to W^{\pm}h/A$
have also been carried out~\cite{Arhrib:2021xmc}.
Detailed overviews of experimental studies of
charged Higgs bosons at the LHC are also
presented in~\cite{Wang:2021pxc,Krab:2022lih,
Arhrib:2024sfg}, including analyses of production via
vectorlike top-quark pairs~\cite{Arhrib:2024nbj} and
studies of the $W\gamma$ decay
mode~\cite{Logan:2018wtm}. Production of charged Higgs
bosons at future lepton colliders has also
been studied at $\mu^+\mu^-$ and $e^+e^-$
machines~\cite{Ouazghour:2023plc,Ouazghour:2024twx,
Ouazghour:2025owf,BrahimAit-Ouazghour:2025mhy}.
Furthermore, heavy charged Higgs states at
$\gamma\gamma$ colliders have been probed using
multivariate analyses including the $H^{\pm}\to
W^{\pm}H$ decay channel~\cite{Ahmed:2024oxg,
Hashemi:2023osd}.

Loop--induced decay $H^{\pm} \to W^{\pm}Z$
is sensitive to BSM effects and provides important
information for discriminating among different
types of THDMs. This decay process was
computed in Ref.~\cite{Arhrib:2006wd}.
Alternative calculations with including the
CP-violating THDM, have been presented in
~\cite{Kanemura:2024ezz,Aiko:2021can,
Hernandez-Sanchez:2004yid,BarradasGuevara:2010xs}.
In this work, one-loop contributions for
decay $H^{\pm} \to W^{\pm}Z$ in the Two-Higgs-Doublet
Model (THDM) are computed and its implications at
future muon--TeV colliders are studied.
Different from other works,
our calculations are performed in the general
$\mathcal{R}_{\xi}$ gauge, and results are verified
through several self-consistency checks such as the
$\xi$-independence, renormalization-scale stability
and ultraviolet finiteness of the amplitude.
A phenomenological analysis is first
carried out in the Type-X THDM by scanning
the updated viable parameter space. Charged Higgs
pair production is then studied via $\mu^+\mu^- \to
H^+H^- \to W^{\pm}W^{\mp}Zh$ and $\mu^+\mu^- \to
\gamma\gamma \to H^+H^- \to W^{\pm}W^{\mp}Zh$,
with signals evaluated with respect to the SM
backgrounds.

Our paper has the following structure.
Reviewing of the THDM framework, its constraints,
and the updated parameter-space scan for
the Type-X THDM are presented in Section~2.
Section~3 presents the one-loop calculation
of $H^{\pm} \to W^{\pm}Z$ along with numerical
checks of the computation. Section~4 discusses
the phenomenological applications of this work.
Section~5 is devoted to the conclusions.
Analytic expressions and checks of
$\xi$-independence are provided in
the appendices.
\section{THDM and its contrainsts} 
A detailed review of the THDM can be found
in Ref.~\cite{Branco:2011iw}.
It is shown that tree-level flavor-changing
neutral currents can be prevented
by applying a discrete $Z_{2}$
symmetry in the Lagrangian, as discussed in
Ref.~\cite{Branco:2011iw}.
The different charge quantum numbers
of the $Z_{2}$ for scalar doublets
and fermion fields lead to
four distinct Yukawa types (known as
Type-I, II, X, Y)
(see also Ref.~\cite{Aoki:2009ha} for more detail).
The Yukawa Lagrangian can be parameterized
as
\begin{eqnarray}
{\mathcal L}_\text{Y}
&=&
-\sum_{f=u,d,\ell}
\left(
\sum_{\phi_j=h, H}
g_{\phi_j ff}\cdot
\phi_j{\overline f}f
+
g_{Aff}\cdot
A
{\overline f}
\gamma_5f
\right)
\\
&&
-
\left[
\bar{u}_{i}
\left(
g_{H^+ u_i d_j}^L m_{u_i}
P_L
+
g_{H^+ u_i d_j}^R
m_{d_j} P_R \right)d_{j} H^+
\right]
\nonumber\\
&&
\nonumber\\
&&
+
\cdots
\nonumber
\\
&=&
-\sum_{f=u,d,\ell}
\left(
\sum_{\phi_j=h, H}
\frac{m_f}{v}\xi_{\phi_j}^f
\phi_j {\overline f}f
-i\frac{m_f}{v}\xi_A^f
{\overline f}
\gamma_5fA
\right)
\\
&&
-
\frac{
\sqrt{2}
}{v}
\left[
\bar{u}_{i}
V_{ij}\left(
m_{u_i}
\xi^{u}_A P_L
+
\xi^{d}_A
m_{d_j} P_R \right)d_{j} H^+
\right]
\nonumber\\
&&
- \frac{\sqrt{2}}{v}
\bar{\nu}_L
\xi^{\ell}_A
m_\ell \ell_R H^+
+ \textrm{H.c}.
\nonumber
\end{eqnarray}
In the Lagrangian, the CKM matrix elements are denoted
by $V_{ij}$, $\ell_{L/R} (\nu_{L/R})$
stand for the left- and right-handed lepton fields,
and $P_{L/R} = (1 \mp
\gamma_{5})/2$ denotes the projection operators.
It is easy to check that the vertices of the charged
Higgs with up- and down-type quarks depend linearly
on $\cot\beta$ in the Type-X THDM. As a result,
fermionic loop contributions are thus diminished
in the large-$t_{\beta}$ regime.

\begin{figure}[H]
\centering
\begin{tabular}{cc}
\includegraphics[width=8cm, height=7cm]
{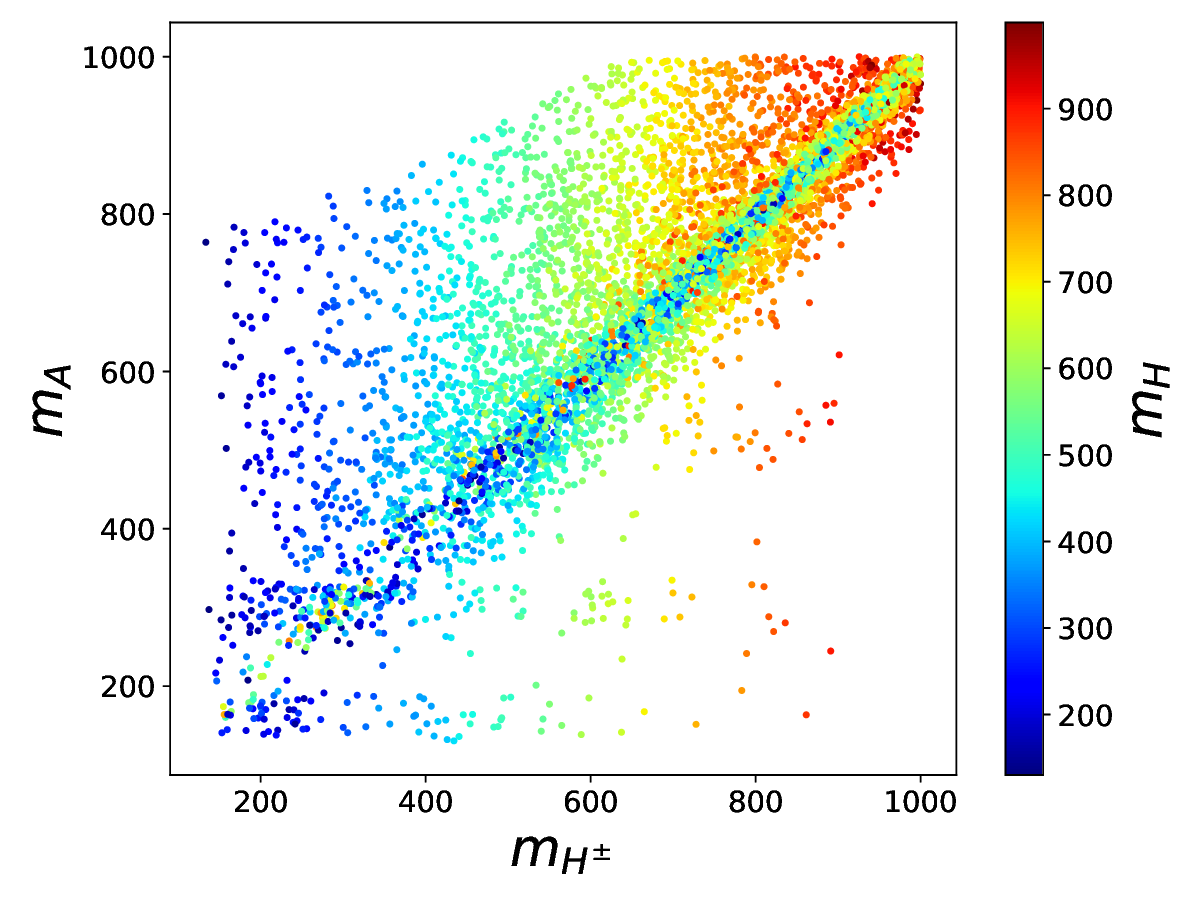}
&
\includegraphics[width=8cm, height=7cm]
{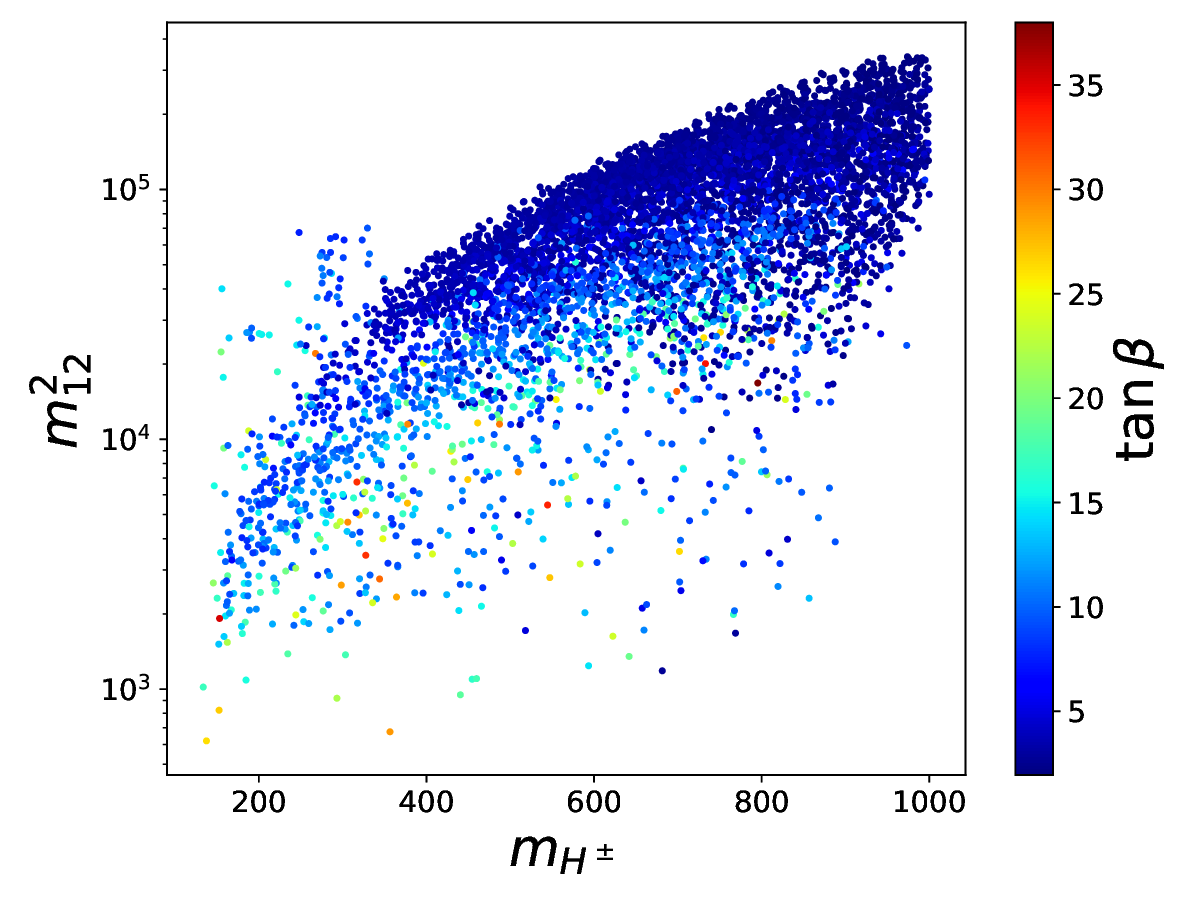}
\end{tabular}
\caption{\label{scanP}
The scatter plots show correlations in the parameter space:
$(m_A,\, m_{H^\pm},\, m_H)$ in the left panel and
$(m_{12}^2,\, m_{H^\pm},\, t_{\beta})$ in the right panel.
}
\end{figure}

We now turn to the theoretical and experimental
bounds on the THDM, which are discussed
in the following paragraphs. Theoretical bounds are
obtained by imposing conditions such as perturbative
unitarity, perturbativity, and vacuum stability, all of
which are taken into account for the models
under consideration.
In the experimental limits, the measured data
of the SM-like Higgs properties, the data
of flavor observables, and electroweak precision
tests are taken into consideration in the
constraints. For explaining these conditions in
detail, we refer to our previous work~\cite{Tran:2025iur}
where Type-I THDM has been studied in
detail. The scan of the parameter space is
performed as follows. We choose parameters
for the Type-X THDM within the ranges
$s_{\beta-\alpha} \in [0.97,1]$,
$t_{\beta} \in [0.5,45]$, $m_{H}
\in [130,1000]~\text{GeV}$,
$m_{A,H^{\pm}} \in [130,1000]~\text{GeV}$,
and $m_{12}^2
\in [0,10^6]~\text{GeV}^2$, with the SM-like
Higgs mass fixed at $m_{h} = 125.09~\text{GeV}$.
The sampling points are first tested against
theoretical constraints. The allowed points are then
checked with the Electroweak Precision Observables (EWPOs).
The surviving parameter space is subsequently passed
to {\tt HiggsBounds-5.10.1}~\cite{Bechtle:2020pkv} and
{\tt HiggsSignals-2.6.1}~\cite{Bechtle:2020uwn}
to incorporate collider
limits and Higgs precision measurement data, respectively.
It is important to stress that both
{\tt HiggsBounds-5.10.1} and
{\tt HiggsSignals-2.6.1} are incorporated
into {\tt 2HDMC}~\cite{Eriksson:2009ws}.
Finally, the remaining points are evaluated with
{\tt SuperIso v4.1}~\cite{Mahmoudi:2008tp} to include flavor
constraints. After all conditions are imposed,
the viable parameter space is thoroughly examined
in the following paragraphs.
In Fig.~\ref{scanP}, the left panel shows the scatter
plot of the viable parameter space in the $(m_A, M_H, m_{H^\pm})$
plane, while the right panel displays the scatter plot in
the $(m_{12}^2, m_{H^\pm}, t_\beta)$ plane.
The results indicate that the
data favor the mass region $m_{A} > m_{H^\pm}=m_H$
over other mass patterns. Across the full
charged Higgs mass range, parameter regions
with $t_{\beta}\leq 10$ and larger $m_{12}^2$
values are preferred, as shown in the
right panel.
\section{One-loop--induced expressions
for $H^{\pm}\rightarrow W^{\pm}Z$ in
the general $R_{\xi}$}
First, all one-loop Feynman diagrams are
generated in the general $R_{\xi}$ gauge
and are shown explicitly in Appendix C.
The decay amplitude for
$H^{\pm}(p) \to W^{\pm}_{\mu}(p_{1}) Z_{\nu}(p_{2})$
can be expressed via the form factors $\mathcal{T}_i$ ($i = 1,2,3$)
following the corresponding Lorentz structures:
\begin{eqnarray}
\mathcal{M}_{H^{\pm}
\rightarrow W^{\pm} Z}
&=&
\Big[
g^{\mu\nu}
\mathcal{T}_1
+
p_2^{\mu}
\,
p_1^{\nu}
\mathcal{T}_2
+
i \,
\epsilon^{\mu\nu\rho\sigma}
p_{1, \rho}
\,
p_{2, \sigma}
\,
\mathcal{T}_3
\Big]
\varepsilon^{*}_{\mu} (p_1)
\varepsilon^{*}_{\nu} (p_2).
\end{eqnarray}
Where $\epsilon^{\mu\nu\rho\sigma}$
is the completely antisymmetric
tensor, $p$ ($p_1$ and $p_2$)
is the ingoing (outgoing) momentum,
and $\varepsilon^{*}_{\mu}$
($\varepsilon^{*}_{\nu}$) are
the polarization vectors for
external $W^\pm$ and $Z$ bosons,
respectively.
In the above formulas,
two relations for
on-shell vector bosons $
p_1^{\mu} 
\varepsilon^{*}_{\mu} (p_1) 
= 
p_2^{\nu} 
\varepsilon^{*}_{\nu} (p_2) 
= 
0$, have been utilized for our
calculations.

The corresponding form factors $\mathcal{T}_i$ are
decomposed into one-loop fermionic ($\mathcal{T}_i^F$)
and bosonic ($\mathcal{T}_i^B$) contributions.
The factors are computed from the 
respective groups of Feynman diagrams
as follows: 
\begin{eqnarray} \mathcal{T}^{(F/B)}_i
&=& \mathcal{T}^{(F/B)}_{i, \text{Trig}}
+ \mathcal{T}^{(F/B)}_{i, \text{Self}}
+ \mathcal{T}^{(F/B)}_{i, \text{Tad}}.
\end{eqnarray} 
The index notations $F/B$ indicate the corresponding
contributions from fermion and boson loops.
The quantities $\mathcal{T}^{(F/B)}_{i, \text{Trig/Self/Tad}}$
are obtained from the triangle, self-energy,
and tadpole Feynman diagrams, respectively.

Analytical results for
$\mathcal{T}^{(F/B)}_{i, \text{Trig/Self/Tad}}$
in the $\mathcal{R}{\xi}$ gauge are presented
using scalar Passarino-Veltman functions
(PV-functions)~\cite{Denner:1991kt} in
appendix~A, while analytical checks
of $\xi$-gauge invariance are provided in appendix~B.
In this section, we perform numerical checks
of the self-consistency of the one-loop form
factors, including their $\xi$-independence,
UV finiteness, and stability under variations
of the renormalization scale $\mu^2$.
Specifically, for
$\xi$-gauge invariance we examine only the form
factors $\mathcal{T}^{B}_{1,2}$ arising from boson-loop
contributions, where $\xi_W$ and $\xi_Z$ are
varied in comparison with the case $\xi_{W/Z} = 1$
in the 't~Hooft--Feynman gauge. For illustration,
we adopt representative THDM parameters: $m_{H^\pm} =
m_{H} = 800~\text{GeV}$, $m_{A} = m_{H^\pm} + m_{Z}$,
$s_{\beta - \alpha} = 0.98$, $t_{\beta} = 10$, and
the scale of the $Z_2$-symmetry
$m_{12}^2 = 5\cdot 10^4~\text{GeV}^2$ (we keep this
parameter point for all numerical checks below).
The results of these checks are summarized in
Tables~\ref{xi-independenceN1},\ref{xi-independenceN2},
where $\xi_W$ and $\xi_Z$ are varied over wide
ranges. The results demonstrate good numerical
stability.
\begin{table}[H]
\centering
\begin{tabular}
{|l|l|l|l|}
\hline
\hline
$\big(
\xi_W
,
\xi_Z
\big)$
&
$(1, 1)$
&
$(10, 10^2)$
&
$(10^3, 10^4)$
\\
\hline \hline
\vspace{-0.325cm}
$\sum
\limits_{\phi}
\mathcal{T}^{B, \phi-A}_{1, \text{Trig}}$
&
$-50.72721116$
&
$-50.79862234$
&
$-51.65760847$
\\
$$
&
$+ 0 \, i$
&
$+ 0 \, i$
&
$+ 0 \, i$
\\ \hline
\vspace{-0.325cm}
$\sum
\limits_{\phi}
\mathcal{T}^{B, \phi-H^\pm}_{1, \text{Trig}}$
&
$50.66247519$
&
$50.66247519$
&
$50.66247519$
\\ 
$$
&
$+ 0 \, i$
&
$+ 0 \, i$
&
$+ 0 \, i$
\\ \hline
\vspace{-0.325cm}
$\sum
\limits_{\phi}
\mathcal{T}^{B, \phi-W^\pm}_{1, \text{Trig}}$
&
$4.197968064$
&
$3.885889625$
&
$9.769973063$
\\ 
$$
&
$-1.192202277 \, i$
&
$-0.7864121884 \, i$
&
$+2.88619913  \, i$
\\ \hline
\vspace{-0.325cm}
$\sum
\limits_{\phi}
\mathcal{T}^{B, \phi-Z}_{1, \text{Trig}}$
&
$0.0173173496$
&
$0.3215636277$
&
$1.42510883$
\\ 
$$
&
$+ 0 \, i$
&
$+ 0 \, i$
&
$+ 0 \, i$
\\ \hline 
\vspace{-0.325cm}
$\sum
\limits_{\phi}
\mathcal{T}^{B, \phi-W^\pm Z}_{1, \text{Trig}}$
&
$-4.072882533$
&
$-3.945505715$
&
$-10.70487047$
\\ 
$$
&
$+1.271771141 \, i$
&
$+0.815048537 \, i$
&
$-3.318528531 \, i$
\\ \hline
$\mathcal{T}^{B}_{1, \text{Self}}$
&
$69.46890566$
&
$78.17143601$
&
$-30.76307909$
\\ 
$$
&
$-0.023317639 \, i$
&
$+0.027614876 \, i$
&
$+0.4885806253 \, i$
\\ \hline
$\mathcal{T}^{B}_{1, \text{Tad}}$
&
$-69.69069235$
&
$-78.44135618$
&
$31.12388118$
\\ 
$$
&
$+ 0 \, i$
&
$+ 0 \, i$
&
$+ 0 \, i$
\\ \hline
$\mathcal{T}^{B}_{1}$
in Eq.~\ref{Eq:NiBTrig}
&
$-0.1441197805$
&
$-0.1441197805$
&
$-0.1441197808$
\\ 
$$
&
$+0.0562512244 \, i$
&
$+0.0562512244 \, i$
&
$+0.0562512244  \, i$
\\
\hline\hline
\end{tabular}
\caption{\label{xi-independenceN1}
Numerical checks of $R_\xi$ gauge invariance
for the form factor $\mathcal{T}^{B}_{1}$ in
boson-loop contributions are performed by varying
the values of $\xi_W$ and $\xi_Z$ and comparing
them with the case $\xi_{W/Z} = 1$ in the
't~Hooft--Feynman gauge. We take the THDM
parameters as follows: the Higgs masses
$m_{H^\pm} = m_{H} = 800$~GeV, $m_{A} = m_{H^\pm} + m_{Z}$,
$s_{\beta - \alpha} = 0.98$, $t_{\beta} = 10$,
and the scale of the $Z_2$-symmetry
$m_{12}^2 = 5 \cdot 10^4$~GeV$^2$.
}
\end{table}
\begin{table}[H]
\centering
\begin{tabular}
{|l|l|l|l|l|}
\hline
\hline
$\big(
\xi_W
,
\xi_Z
\big)$
&
$(1, 1)$
&
$(10, 10^2)$
&
$(10^3, 10^4)$
\\
\hline \hline
\vspace{-0.325cm}
$\sum
\limits_{\phi}
\mathcal{T}^{B, \phi-A}_{2, \text{Trig}}$
&
$8.875933961
\cdot 10^{-7}$
&
$8.875933961
\cdot 10^{-7}$
&
$8.875933958
\cdot 10^{-7}$
\\ 
$$
&
$+ 0 \, i$
&
$+ 0 \, i$
&
$+ 0 \, i$
\\ \hline
\vspace{-0.325cm}
$\sum
\limits_{\phi}
\mathcal{T}^{B, \phi-H^\pm}_{2, \text{Trig}}$
&
$-4.259326679
\cdot 10^{-7}$
&
$-4.259326679
\cdot 10^{-7}$
&
$-4.259326683
\cdot 10^{-7}$
\\ 
$$
&
$+ 0 \, i$
&
$+ 0 \, i$
&
$+ 0 \, i$
\\ \hline
\vspace{-0.325cm}
$\sum
\limits_{\phi}
\mathcal{T}^{B, \phi-W^\pm}_{2, \text{Trig}}$
&
$3.525339605
\cdot 10^{-7}$
&
$3.525339607
\cdot 10^{-7}$
&
$3.52533961
\cdot 10^{-7}$
\\ 
$$
&
$+5.18767908
\cdot 10^{-7} \, i$
&
$+5.18767908
\cdot 10^{-7} \, i$
&
$+5.1876791
\cdot 10^{-7} \, i$
\\ \hline
\vspace{-0.325cm}
$\sum
\limits_{\phi}
\mathcal{T}^{B, \phi-Z}_{2, \text{Trig}}$
&
$-7.0029171
\cdot 10^{-8}$
&
$-7.0029171
\cdot 10^{-7}$
&
$-7.0029174
\cdot 10^{-7}$
\\ 
$$
&
$+ 0 \, i$
&
$+ 0 \, i$
&
$+ 0 \, i$
\\ \hline 
\vspace{-0.325cm}
$\sum
\limits_{\phi}
\mathcal{T}^{B, \phi-W^\pm Z}_{2, \text{Trig}}$
&
$-8.244771
\cdot 10^{-7}$
&
$-8.244771
\cdot 10^{-7}$
&
$-8.244773
\cdot 10^{-7}$
\\ 
&
$-4.926311
\cdot 10^{-7} \, i$
&
$-4.926311
\cdot 10^{-7} \, i$
&
$-4.926313
\cdot 10^{-7} \, i$
\\ \hline
$\mathcal{T}^{B}_{2, \text{Self}}$
&
$0$
&
$0$
&
$0$
\\ \hline
$\mathcal{T}^{B}_{2, \text{Tad}}$
&
$0$
&
$0$
&
$0$
\\ \hline
$\mathcal{T}^{B}_{2}$
in Eq.~\ref{Eq:NiBTrig}
&
$-8.031155
\cdot 10^{-8}$
&
$-8.031155
\cdot 10^{-8}$
&
$-8.031154
\cdot 10^{-8}$
\\ 
&
$+2.6136808
\cdot 10^{-8} \, i$
&
$+2.6136809
\cdot 10^{-8} \, i$
&
$+2.613682
\cdot 10^{-8} \, i$
\\
\hline\hline
\end{tabular}
\caption{\label{xi-independenceN2}
Numerical checks of $R_\xi$ gauge invariance
for the form factor $\mathcal{T}^{B}_{2}$ in
boson-loop contributions are performed by varying
the gauge parameters $\xi_W$ and $\xi_Z$. We take
the THDM parameters as follows: the Higgs masses
$m_{H^\pm} = m_{H} = 800$~GeV,
$m_{A} = m_{H^\pm} + m_{Z}$,
$s_{\beta - \alpha} = 0.98$, $t_{\beta} = 10$, and
the scale of the $Z_2$-symmetry
$m_{12}^2 = 5 \times 10^4$~GeV$^2$.
}
\end{table}
We then perform numerical checks of the
$C_{UV}$- and $\mu^2$-independence
for the form factors $\mathcal{T}_i = \mathcal{T}_i^F
+ \mathcal{T}_i^B$ with $i=1,2,3$. We note
that the total one-loop form factor should
be considered in these tests, with gauge
parameters fixed at $\xi_W = \xi_Z = 100$
for example. It should also be noted
that the fermion-loop contribution
$\mathcal{T}^{F}_{1}$ is evaluated in the
Type-X THDM as an illustrative example.
The numerical results for these tests are
obtained using the same parameter point as
specified above. By varying $C_{UV}$ and
$\mu^2$ over wide ranges, the results
demonstrate good numerical stability (see
Tables~\ref{cuv1}, \ref{cuv2}, \ref{cuv3}).
\begin{table}[H]
\centering
\begin{tabular}
{|l|l|l|l|}
\hline\hline
$\big(
C_{UV}
,
\mu^2
\big)$
&
$(0,1)$
&
$(10^4, 10^6)$
&
$(10^6, 10^8)$
\\
\hline \hline
$\mathcal{T}^{B}_{1, \text{Trig}}$
&
$0.0409007639$
&
$0.0409007639$
&
$0.04090076301$
\\ 
&
$-0.12057752761 \, i$
&
$-0.12057752761 \, i$
&
$-0.12057752761 \, i$
\\ \hline
$\mathcal{T}^{B}_{1, \text{Self}}$
&
$-0.9723666455$
&
$799.5362885$
&
$79940.92392$
\\ 
&
$+0.1590201866 \, i$
&
$+0.1590201866 \, i$
&
$+0.1590201866 \, i$
\\ \hline
$\mathcal{T}^{B}_{1, \text{Tad}}$
&
$1.0786977518$
&
$-799.4299574$
&
$-79940.81759$
\\ 
&
$+ 0 \, i$
&
$+ 0 \, i$
&
$+ 0 \, i$
\\ \hline
$\mathcal{T}^{B}_{1}$
&
$0.1472318702$
&
$0.1472318702$
&
$0.1472318695$
\\ 
&
$+0.03844265898 \, i$
&
$+0.03844265898 \, i$
&
$+0.03844265898 \, i$
\\ \hline 
$\mathcal{T}^{F}_{1, \text{Trig}}$
&
$0.4996708114$
&
$-497.4139502$
&
$-49723.08395$
\\ 
&
$-0.19005144066 \, i$
&
$-0.19005144066 \, i$
&
$-0.19005144066 \, i$
\\ \hline
$\mathcal{T}^{F}_{1, \text{Self}}$
&
$-0.3824890675$
&
$375.5060921$
&
$37537.3078$
\\ 
& 
$+0.12113853706 \, i$
&
$+0.12113853706 \, i$
&
$+0.12113853706 \, i$
\\ \hline
$\mathcal{T}^{F}_{1, \text{Tad}}$
&
$-0.1133787187$
&
$121.9116612$
&
$12185.77995$
\\ 
&
$+ 0 \, i$
&
$+ 0 \, i$
&
$+ 0 \, i$
\\ \hline
$\mathcal{T}^{F}_{1}$
&
$0.00380302517$
&
$0.00380302517$
&
$0.00380302519$
\\ 
&
$-0.06891290359 \, i$
&
$-0.06891290359 \, i$
&
$-0.06891290359 \, i$
\\ \hline 
$\mathcal{T}_1=\mathcal{T}_1^F + \mathcal{T}_1^B$
&
$0.1510348954$
&
$0.1510348954$
&
$0.1510348947$
\\ 
&
$-0.0304702446 \, i$
&
$-0.0304702446 \, i$
&
$-0.0304702446 \, i$
\\
\hline
\hline
\end{tabular}
\caption{\label{cuv1}
Numerical checks of $C_{UV}$ and the
renormalization scale $\mu^2$ are performed
for the form factors
$\mathcal{T}_1 = \mathcal{T}_1^F + \mathcal{T}_1^B$.
The bosonic contribution $\mathcal{T}^{B}_{1}$ is
evaluated at $\xi_W = \xi_Z = 100$,
while the fermionic contribution $\mathcal{T}^{F}_{1}$
is calculated in the Type-X THDM.
For this analysis, we adopt the following
set of THDM parameters:
$ m_{H^\pm} = m_{H} = 500~\text{GeV},\;
m_{A} = m_{H^\pm} + m_{Z}, \;
s_{\beta - \alpha} = 0.98, \;
t_{\beta} = 5, \;
m_{12}^2 = 5 \times 10^4~\text{GeV}^2$.
}
\end{table}

\begin{table}[H]
\centering
\begin{tabular}
{|l|l|l|l|}
\hline\hline
$\big(
C_{UV}
,
\mu^2
\big)$
&
$(0,1)$
&
$(10^4, 10^6)$
&
$(10^6, 10^8)$
\\
\hline 
\hline
$\mathcal{T}^{B}_{2, \text{Self}}$
&
$0$
&
$0$
&
$0$
\\ \hline
$\mathcal{T}^{B}_{2, \text{Tad}}$
&
$0$
&
$0$
&
$0$
\\ \hline
$\mathcal{T}^{B}_{2}
\equiv
\mathcal{T}^{B}_{2, \text{Trig}}$
&
$-4.46523589
\cdot 10^{-7}$
&
$-4.46523589
\cdot 10^{-7}$
&
$-4.465235912
\cdot 10^{-7}$
\\ 
&
$+8.283757782
\cdot 10^{-8} \, i$
&
$+8.283757782
\cdot 10^{-8} \, i$
&
$+8.283757782
\cdot 10^{-8} \, i$
\\ \hline 
$\mathcal{T}^{F}_{2, \text{Self}}$
&
$0$
&
$0$
&
$0$
\\ \hline
$\mathcal{T}^{F}_{2, \text{Tad}}$
&
$0$
&
$0$
&
$0$
\\ \hline
$\mathcal{T}^{F}_{2}$
&
$1.039941873
\cdot 10^{-7}$
&
$1.039941873
\cdot 10^{-7}$
&
$1.039941873
\cdot 10^{-7}$
\\ 
&
$+1.445658912
\cdot 10^{-6} \, i$
&
$+1.445658912
\cdot 10^{-6} \, i$
&
$+1.445658912
\cdot 10^{-6} \, i$
\\ \hline 
\hline
$\mathcal{T}_2=\mathcal{T}_2^F + \mathcal{T}_2^B$
&
$-3.425294017
\cdot 10^{-7}$
&
$-3.425294017
\cdot 10^{-7}$
&
$-3.425294039
\cdot 10^{-7}$
\\ 
&
$+1.52849649
\cdot 10^{-6} \, i$
&
$+1.52849649
\cdot 10^{-6} \, i$
&
$+1.52849649
\cdot 10^{-6} \, i$
\\
\hline
\hline
\end{tabular}
\caption{\label{cuv2}
Numerical checks of $C_{UV}$ and the
renormalization scale $\mu^2$
are performed for the form factors
$\mathcal{T}_2 = \mathcal{T}_2^F + \mathcal{T}_2^B$.
The bosonic contribution $\mathcal{T}^{B}_{2}$ is evaluated at
$\xi_W = \xi_Z = 100$, while the fermionic contribution
$\mathcal{T}^{F}_{2}$ is calculated in the Type-X THDM.
For this analysis, we adopt the following THDM parameters:
$m_{H^\pm} = m_{H} = 500~\text{GeV}$, $m_{A} = m_{H^\pm} + m_{Z}$,
$s_{\beta - \alpha} = 0.98$, $t_{\beta} = 5$, and
$m_{12}^2 = 5 \times 10^4~\text{GeV}^2$.
}
\end{table}
\begin{table}[H]
\centering
\begin{tabular}
{|l|l|l|l|}
\hline\hline
$\big(
C_{UV}
,
\mu^2
\big)$
&
$(0,1)$
&
$(10^4, 10^6)$
&
$(10^6, 10^8)$
\\
\hline 
\hline
$\mathcal{T}^{B}_{3}$
&
$0$
&
$0$
&
$0$
\\ \hline 
$\mathcal{T}^{F}_{3, \text{Self}}$
&
$0$
&
$0$
&
$0$
\\ \hline
$\mathcal{T}^{F}_{3, \text{Tad}}$
&
$0$
&
$0$
&
$0$
\\ \hline
$\mathcal{T}^{F}_{3}$
&
$3.220832776
\cdot 10^{-7}$
&
$3.220832776
\cdot 10^{-7}$
&
$3.220832777
\cdot 10^{-7}$
\\ 
&
$-1.274849257
\cdot 10^{-6} \, i$
&
$-1.274849257
\cdot 10^{-6} \, i$
&
$-1.274849257
\cdot 10^{-6} \, i$
\\ \hline 
$\mathcal{T}_3= \mathcal{T}_3^F + \mathcal{T}_3^B$
&
$3.220832776
\cdot 10^{-7}$
&
$3.220832776
\cdot 10^{-7}$
&
$3.220832777
\cdot 10^{-7}$
\\ 
&
$-1.274849257
\cdot 10^{-6} \, i$
&
$-1.274849257
\cdot 10^{-6} \, i$
&
$-1.274849257
\cdot 10^{-6} \, i$
\\
\hline
\hline
\end{tabular}
\caption{
\label{cuv3}
Numerical checks of $C_{UV}$ and
the renormalization scale $\mu^2$
are performed for the form factors
$\mathcal{T}_3 = \mathcal{T}_3^F
+ \mathcal{T}_3^B$.
The bosonic contribution
$\mathcal{T}^{B}_{3}$ is evaluated at
$\xi_W = \xi_Z = 100$, while the
fermionic contribution
$\mathcal{T}^{F}_{3}$ is calculated
in the Type-X THDM as an illustrative
example of fermion couplings. For
this analysis, we adopt the following
THDM parameters: $m_{H^\pm} = m_{H} = 500~\text{GeV}$,
$m_{A} = m_{H^\pm} + m_{Z}$, $s_{\beta - \alpha} = 0.98$,
$t_{\beta} = 5$, and $m_{12}^2
= 5 \times 10^4~\text{GeV}^2$.
}
\end{table}
After collecting all the necessary one-loop
form factors and performing the self-consistency
checks, the decay rates are computed
in terms of these form factors
as follows:
\begin{eqnarray}
\Gamma_{H^{\pm}\rightarrow
W^{\pm} Z}
&=&
\frac{
\sqrt{
\Lambda
(\mu_{W},
\mu_{Z})
}
}{
128 \pi
\cdot
m_{H^\pm}}
\Big\{
4
\big|
\mathcal{T}_1
\big|^2
+
m_{H^\pm}^4
\,
\Lambda
(\mu_{W},
\mu_{Z})
\,
\big|
\mathcal{T}_3
\big|^2
\nonumber\\
&&
+
\frac{
m_{H^\pm}^4
}{
16 m_{W}^2
m_{Z}^2}
\Big|
2
\big(
1 - \mu_{W} - \mu_{Z}
\big)
\,
\mathcal{T}_1
+
m_{H^\pm}^2
\,
\Lambda
(\mu_{W},
\mu_{Z})
\,
\mathcal{T}_2
\Big|^2
\Big\}.
\end{eqnarray}
The relevant kinematical
variables are
$\mu_{V} = m_{V}^2 / m_{H^\pm}^2$
for $V = W, Z$, and the kinematical
function $\Lambda(x, y)$
is defined as follows
$\Lambda(x, y) = (1 - x - y)^2 - 4 x y$.
\section{Production of Singly Charged
Higgs Bosons at muon--TeV Colliders }
Singly charged Higgs boson production
at muon--TeV colliders
is investigated in this section. For the phenomenological
analysis, we use the following benchmark
configuration: $s_{\beta-\alpha} = 0.98$, 
$m_{A} = m_{H^\pm}+m_Z = m_H + m_Z$, 
and $m_{12}^2 = 5\cdot 10^{4}~\text{GeV}^2$.
The charged Higgs mass is scanned over the range
$200~\text{GeV} \leq m_{H^\pm} \leq 1000~\text{GeV}$, 
while the mixing parameter is chosen as 
$2 \leq t_{\beta} \leq 12$. All other SM 
parameters are taken from the Particle 
Data Group~\cite{ParticleDataGroup:2024cfk}.  
\subsection{Branching fractions}
\begin{figure}[H]
\centering
\begin{tabular}{cc}
\hspace{-4cm}
Br$\{H^{\pm} \rightarrow XY\}$
&
\hspace{-4cm}
Br$\{H^{\pm} \rightarrow XY\}$
\\
\includegraphics[width=8cm, height=7cm]
{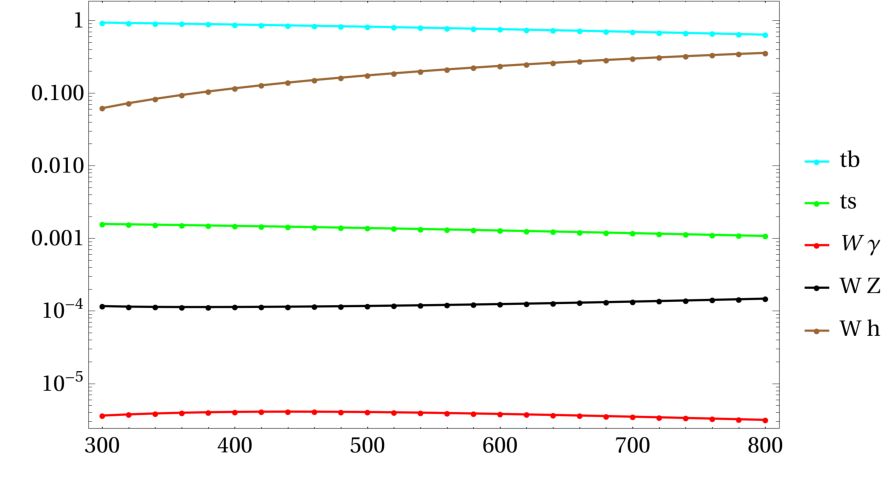}
 &
\includegraphics[width=8cm, height=7cm]
{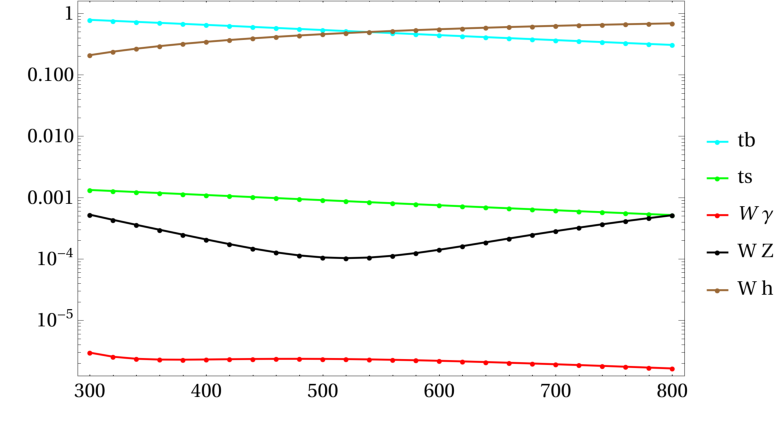}
\\
\includegraphics[width=8cm, height=7cm]
{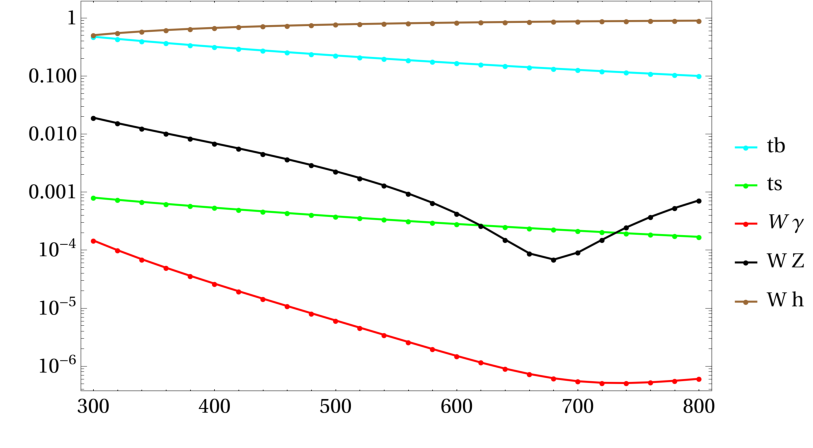}
 &
\includegraphics[width=8cm, height=7cm]
{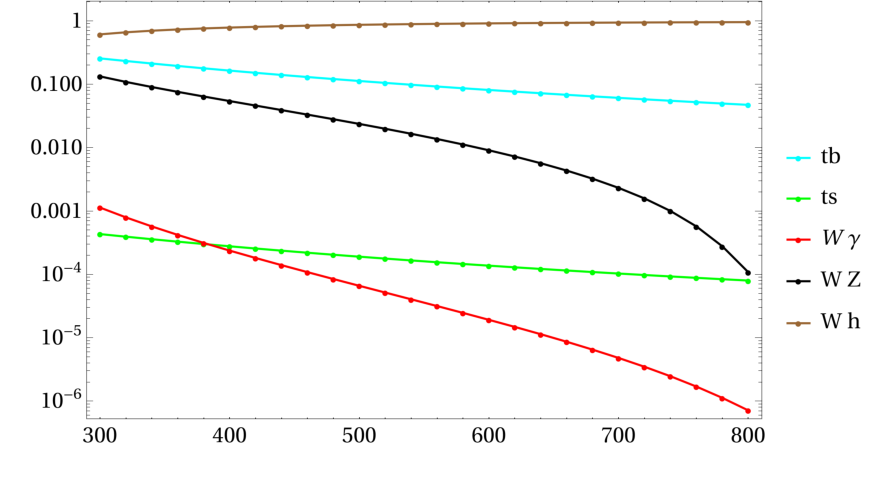}
\\
\hspace{4cm}
$m_{H^\pm}$ [GeV]
&
\hspace{4cm}
$m_{H^\pm}$ [GeV]
\end{tabular}
\caption{\label{BranchingHpm1}
The charged Higgs decay branching ratios in
the Type-X THDM are shown for the benchmark
scenario described above.
Each plot is shown for $t_{\beta} = 2$ (top left),
$t_{\beta} = 4$ (top right), $t_{\beta} = 8$ (bottom left),
and $t_{\beta} = 12$ (bottom right).
}
\end{figure}

We first evaluate the branching fractions of
the charged Higgs boson in the Type-X THDM.
The one-loop--induced process
$H^{\pm} \rightarrow W^\pm \gamma$
has already been reported in our previous
work~\cite{Tran:2025iur} while
decay mode $H^{\pm} \rightarrow W^\pm Z$
is devoted in this work. The remaining
decay channels are taken from~\cite{Aiko:2021can}.
In Fig.~\ref{BranchingHpm1}, the branching
fractions of the charged Higgs boson are
shown for all considered decay modes
within the interval $200~\text{GeV}
\leq m_{H^\pm} \leq 1000~\text{GeV}$.
At $t_{\beta}=2$,
the $tb$ and $Wh$ channels
are the leading contributions, whereas the
branching ratio of $H^{\pm} \rightarrow
W^\pm Z$ remains of the order $10^{-4}$
across the entire mass range. For
$t_{\beta}=4$, the $tb$ and $Wh$ channels
continue to dominate, while the $H^{\pm}
\rightarrow W^\pm Z$ mode stays around
$\sim 10^{-4}$. Next, we consider
the branching fractions at
$t_{\beta}=8$. We find that the branching
ratio of $H^{\pm} \rightarrow W^\pm Z$
increases to the order of $10^{-2}$ in the
low-mass region but decreases to about
$10^{-3}$ at higher masses. Finally, we
examine the case of $t_{\beta}=12$. The
results show that the $H^{\pm} \rightarrow
W^\pm Z$ branching ratio can reach the order
of $10^{-1}$ in the low-mass region, while
it remains at the level of $10^{-4}$ for
larger charged Higgs masses.
As indicated in the previous section, the
fermionic-loop contributions
are suppressed in the
high-$t_{\beta}$ regime. The interference
between the fermionic-loop and bosonic-loop
contributions is small and has the opposite
sign compared with the squared bosonic
contributions. As a result, this leads to
the enhanced decay rates of the $H^{\pm}
\rightarrow W^\pm Z$ mode in the
high-$t_{\beta}$ region compared
with the small-$t_{\beta}$ regime.
\subsection{Processes
$\mu^+\mu^- \rightarrow H^{+}H^{-}
\rightarrow W^{\pm}W^{\mp}Zh$}
\begin{figure}[H]
\centering
\begin{tabular}{cc}
\includegraphics[width=8cm,height=8cm]
{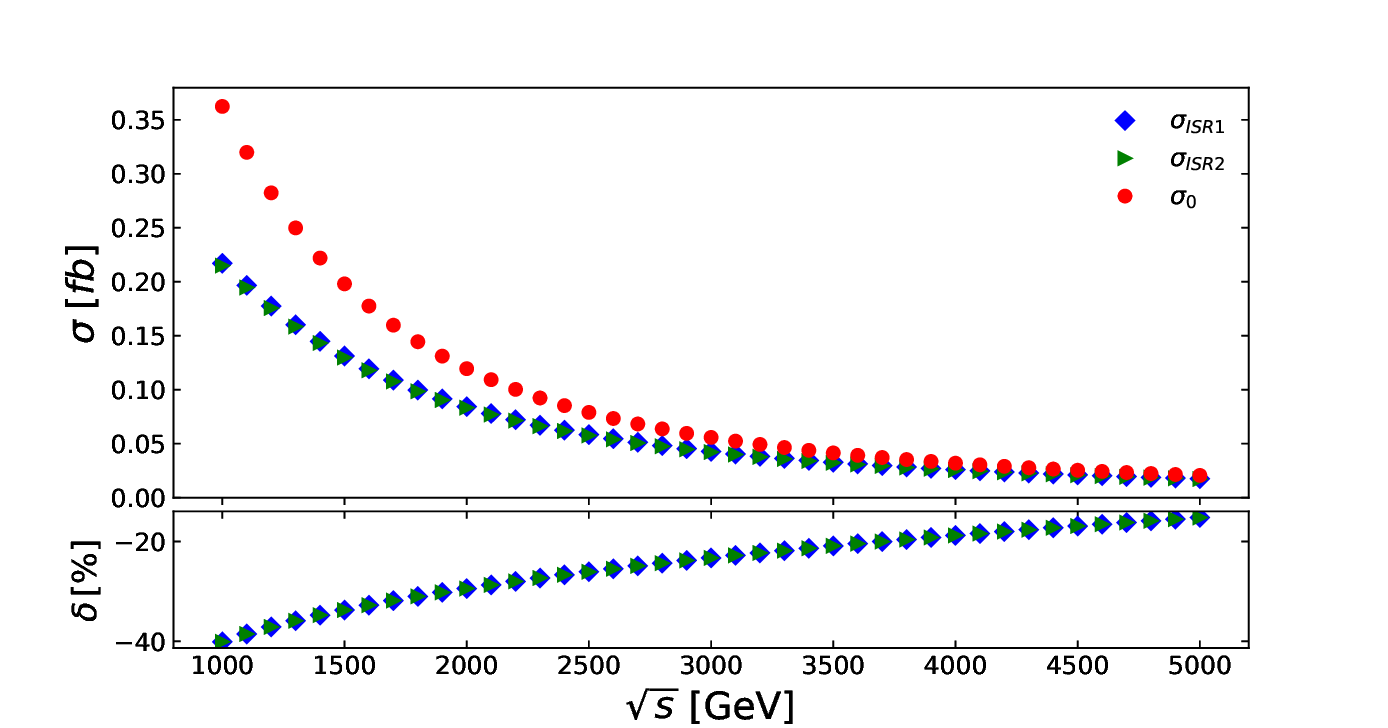}
&
\includegraphics[width=8cm,height=8cm]
{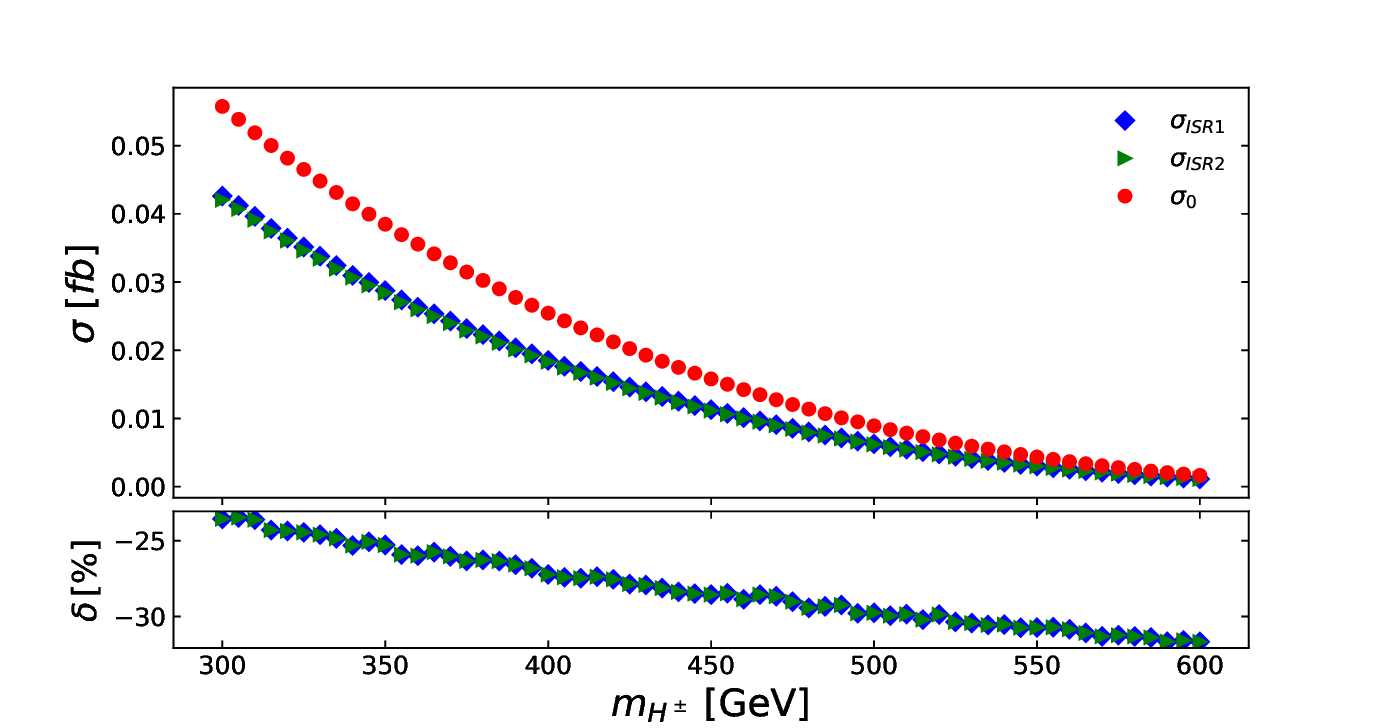}
\end{tabular}
\caption{\label{ISRex}
The effects of ISR on the scattering process 
$\mu^+\mu^- \rightarrow H^{+}H^{-} 
\rightarrow W^{\pm}W^{\mp}Zh$ are examined as 
functions of the center-of-mass energy 
(left panel) and of the charged Higgs mass
at $\sqrt{s}=3$~TeV (right panel).}
\end{figure}

\begin{figure}[H]
\centering
\begin{tabular}{cc}
\includegraphics[width=8cm, height=9cm]
{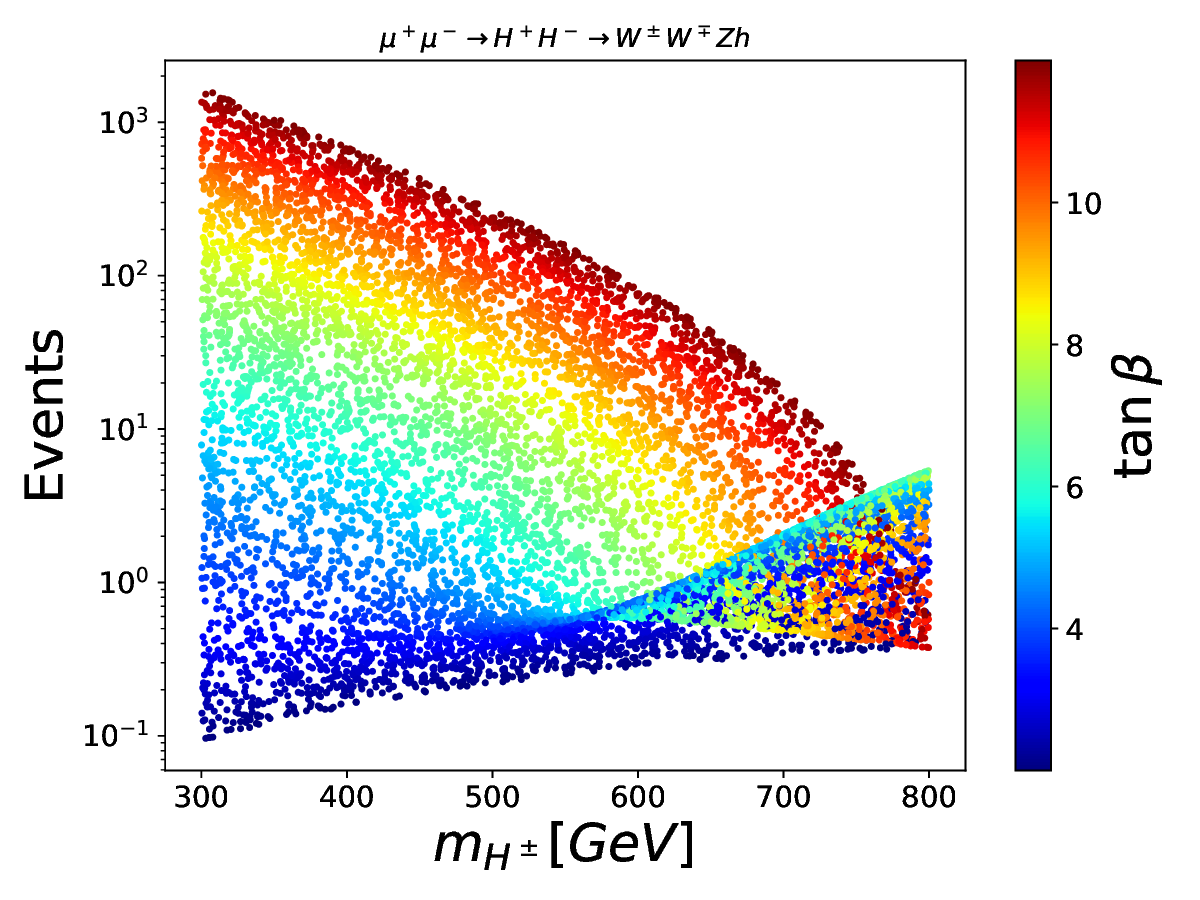}
&
\includegraphics[width=7cm, height=9cm]
{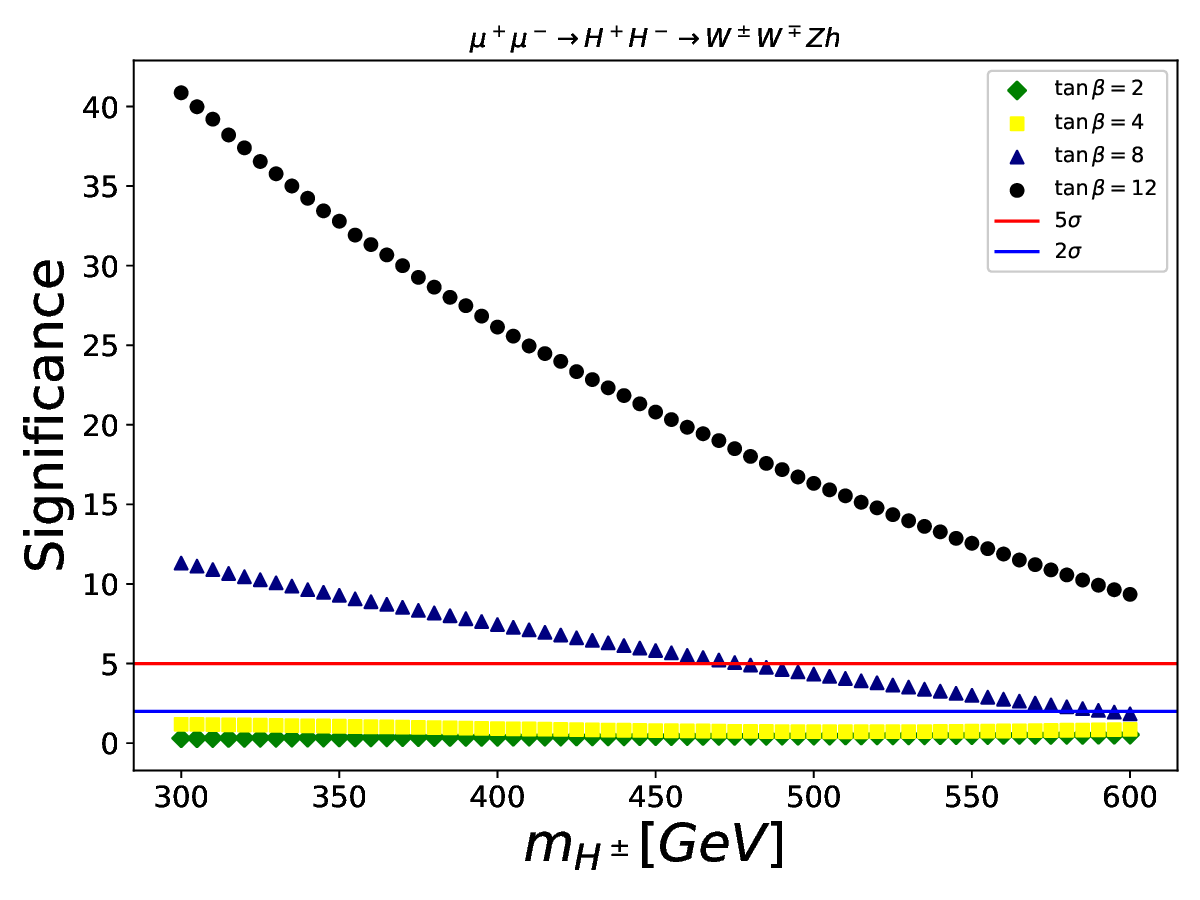}
\end{tabular}
\caption{\label{event1}
The event distributions of the process
$\mu^+\mu^- \rightarrow H^{+}H^{-} \rightarrow W^{\pm}W^{\mp}Zh$
at an integrated luminosity of $\mathcal{L}=3000~\text{fb}^{-1}$,
including ISR corrections, are shown in the left panel.
The corresponding signal significance taking into account
the Standard Model backgrounds, is presented in the right panel.}
\end{figure}

We investigate the potential to probe charged Higgs
pair production by analyzing the process
$\mu^+\mu^- \rightarrow H^{+}H^{-} \rightarrow
W^{\pm}W^{\mp}Zh$ at muon--TeV colliders.
It is well known that
initial-state radiation (ISR) effects play a
crucial role at future lepton colliders. These
effects must be taken into account when
simulating the signals of the charged Higgs.
By applying the factorization theorems for soft
and collinear singularities, the ISR
contributions to charged Higgs pair production
can be calculated as the
following master formula:
\begin{eqnarray}
\label{ISRmaster}
d\sigma (s) = \int d x_1 d x_2
D(x_1, s) D(x_2, s) d\sigma_0 (x_1 x_2 s)
\Theta (\textrm{cuts}).
\end{eqnarray}
In Eq.~(\ref{ISRmaster}), $d\sigma_0$ denotes for
tree-level differential cross sections for both
signals $\mu^-\mu^+ \to H^\pm H^\mp \to
W^{\pm}W^{\mp}Zh$ and the SM background
process $\mu^-\mu^+ \to
W^{\pm}W^{\mp}Zh$.
Additionally, $\Theta (\textrm{cuts})$
represents the appropriate cuts applied in the
simulation, as shown explicitly in the following
paragraphs. $D(x_2, s)$ is the structure function
(SF). The explicit expressions for the SF functions
are presented in the following paragraphs.
The all-order SF functions, which are valid
in the soft-photon limit, are given by:
\begin{eqnarray}
\label{eq:DGL}
D_{GL} (x,s ) =
\frac{\exp\left[  \frac{1}{2}
\beta \left( \frac{3}{4} - \gamma_E \right)
\right]}{\Gamma
\left( 1 + \frac{1}{2} \beta \right) }
\frac{1}{2} \beta
\left( 1 - x \right)^{\frac{1}{2} \beta - 1},
\end{eqnarray}
where
\begin{equation}
\beta =  \frac{2 \alpha}{\pi} (L - 1),
\quad L = \ln(s/m_{\ell}^2).
\end{equation}
Here, $\alpha$ is the fine-structure constant,
and $m_{\ell}$ denotes the lepton mass. The symbol
$\Gamma$ represents the Gamma function,
and $\gamma_E$ is the Euler–Mascheroni constant.
Photon radiation can be treated in the collinear
approximation and through collinear logarithmic
enhancements, which are included in the
large-$\beta$ factor. According to Eq.~(\ref{eq:DGL}),
the additive SF function up to third-order expansion
terms is expressed as follows~\cite{Cacciari:1992pz}:
\begin{eqnarray}
\label{DA}
D_A (x,s) &=& 
D_{GL} (x,s) 
- \frac{1}{4} \beta (1+x) 
\\
&&
+
\frac{1}{32} \beta^2 
\left[ \left(1+x \right) 
\left(-4 \ln (1-x) + 3 \ln(x) \right)
-4 \frac{\ln x}{1-x} -5 -x \right] 
\nonumber\\
&&
+
\frac{1}{384} \beta^3 \Bigg\{ (1+x) 
\left[ 18 \zeta(2) -6 {\rm Li}_2 (x)
-12 \ln^2 (1-x) \right]  \nonumber \\
&& + \frac{1}{1-x} 
\left[ - \frac{3}{2} (1 + 8 x 
+ 3 x^2) \ln x -6 (x+5) (1-x) 
\ln(1-x) \right.  \nonumber \\
&& -12 (1+x^2) \ln x \ln(1-x) 
+ \frac{1}{2} (1+7 x^2) \ln^2 x
- 
\left.  
\frac{1}{4} (39 - 24 x - 15 x^2) 
\right] \Bigg\}.
\nonumber
\end{eqnarray}
In this structure function,
the Riemann $\zeta$ function is
taken into account and ${\rm Li}_2$ is
the dilogarithm functions.
Furthermore, the factorized SF function
up to third order expansion terms
can be obtained
as~\cite{Skrzypek:1990qs,Skrzypek:1992vk}:
\begin{eqnarray}
\label{DF}
D_F (x,s) &=& D_{GL} (x,s) 
\times 
\\
&&
\times
\Bigg\{ 
\frac{1}{2} (1 + x^2) 
-
\frac{\beta}{16} 
\left[ (1 + 3 x^2) \ln x 
+ 2(1-x)^2 \right]  
\nonumber\\
&&
+
\frac{\beta^2}{32} 
\Big[ (1-x)^2 + \frac{1}{2} 
(3x^2 - 4x+1) \ln x 
\nonumber\\
&&
+ \frac{1}{12} (1+7x^2) \ln^2 x 
+ (1-x^2) {\rm Li}_2 (1-x) \Big] 
\Bigg\}.
\nonumber
\end{eqnarray}
In this work, we consider the ISR effects for
both the signal and the SM background processes.
The effects of ISR on the scattering process
$\mu^+\mu^- \rightarrow H^{+}H^{-} \rightarrow
W^{\pm}W^{\mp}Zh$ are examined as functions of
the center-of-mass energy (left panel) and of
the charged Higgs mass at $\sqrt{s}=3$~TeV
(right panel). In the center-of-mass energy range
from $1000$~GeV to $5000$~GeV, ISR corrections
vary from $-40\%$ to $-20\%$, approximately. For
the right-panel plot, we study ISR corrections at
$\sqrt{s}=3000$~GeV while varying the charged
Higgs masses. The results indicate that
the corrections are from $-20\%$ to $-30\%$
for charged Higgs masses
in the interval $[300, 600]$~GeV. In
Fig.~\ref{ISRex}, $\sigma_{ISR1}$ ($\sigma_{ISR2}$)
denotes the cross section calculated using the
structure functions from Eqs.~\ref{DA} and~\ref{DF}.
Using the cross section with ISR corrections,
we evaluate the signal events for
$\mu^+\mu^- \rightarrow H^{+}H^{-} \rightarrow
W^{\pm}W^{\mp}Zh$ at $\sqrt{s}=3$~TeV and the
integrated luminosity of $3000$~fb$^{-1}$. The
events are generated in the parameter space of
$m_{H^\pm}$ and the mixing angle $t_{\beta}$
as shown in Fig.~\ref{event1}. In
this study, we vary $300~\text{GeV} \leq
m_{H^\pm} \leq 600~\text{GeV}$ and $2 \leq
t_{\beta} \leq 12$. The results indicate that the
signal events are significant in regions of low
charged Higgs masses and large $t_{\beta}$
values, while in other regions the events become
negligible. In the right-panel plot, the
significance of the signals relative to the SM
backgrounds is presented. The SM background is
calculated using the {\tt GRACE}
program~\cite{Belanger:2003sd}.
To reduce the SM background, we apply cuts on the
invariant masses of the final-state particles:
$|m_{Wh}-m_{H^\pm}| < 10$~GeV and
$|m_{WZ}-m_{H^\pm}| < 10$~GeV. The significances
are shown for $t_{\beta}=2$ (green points),
$t_{\beta}=4$ (yellow points), $t_{\beta}=8$
(blue points), and $t_{\beta}=12$ (black points).
Our results indicate that in the low-mass regions
of the charged Higgs, and for $t_{\beta}=8$ and
$10$, the significances can exceed $5\sigma$, while
in other regions they become negligible.
\subsection{Processes
$\mu^+\mu^- \rightarrow \gamma \gamma
\rightarrow H^{+}H^{-}
\rightarrow W^{\pm}W^{\mp}Zh$}
\begin{figure}[H]
\centering
\begin{tabular}{cc}
\includegraphics[width=8cm, height=8cm]
{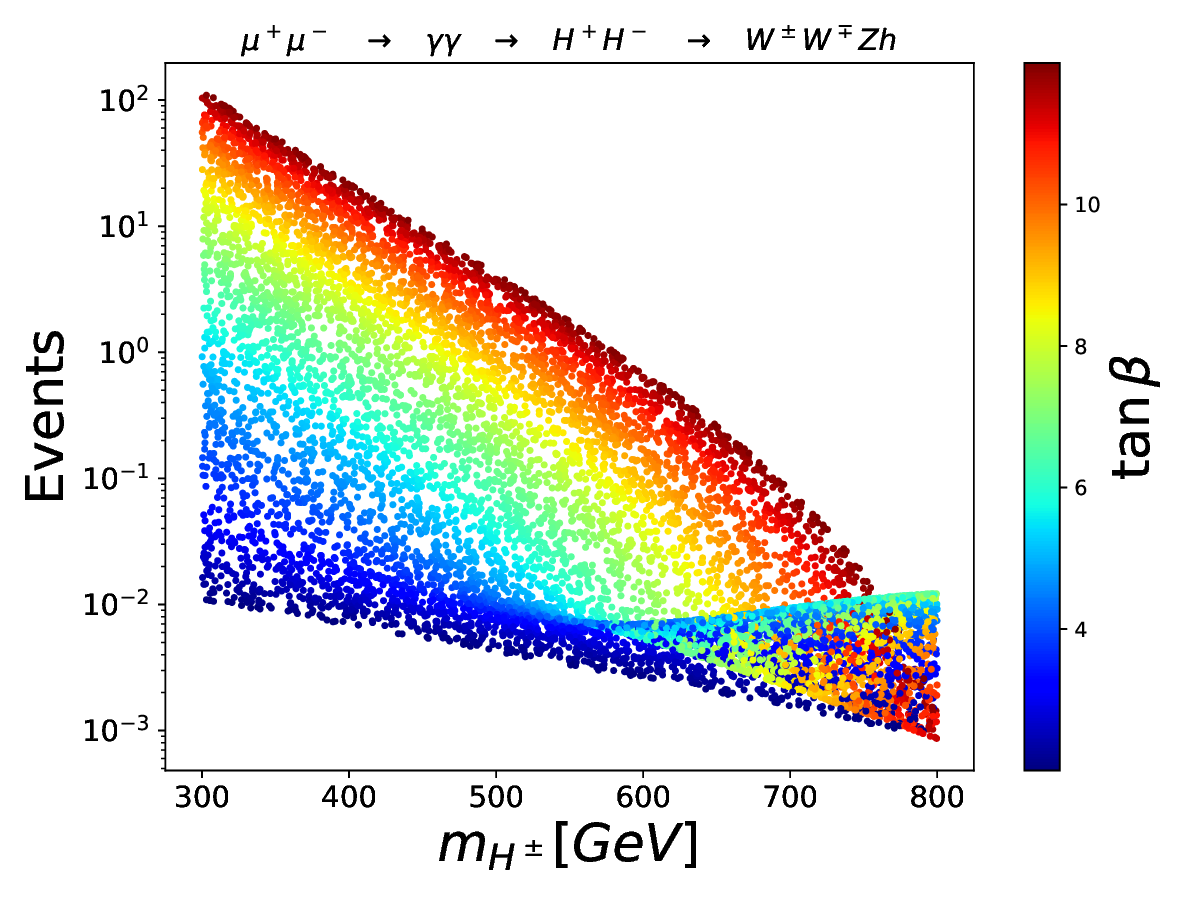}
&
\includegraphics[width=8cm, height=8cm]
{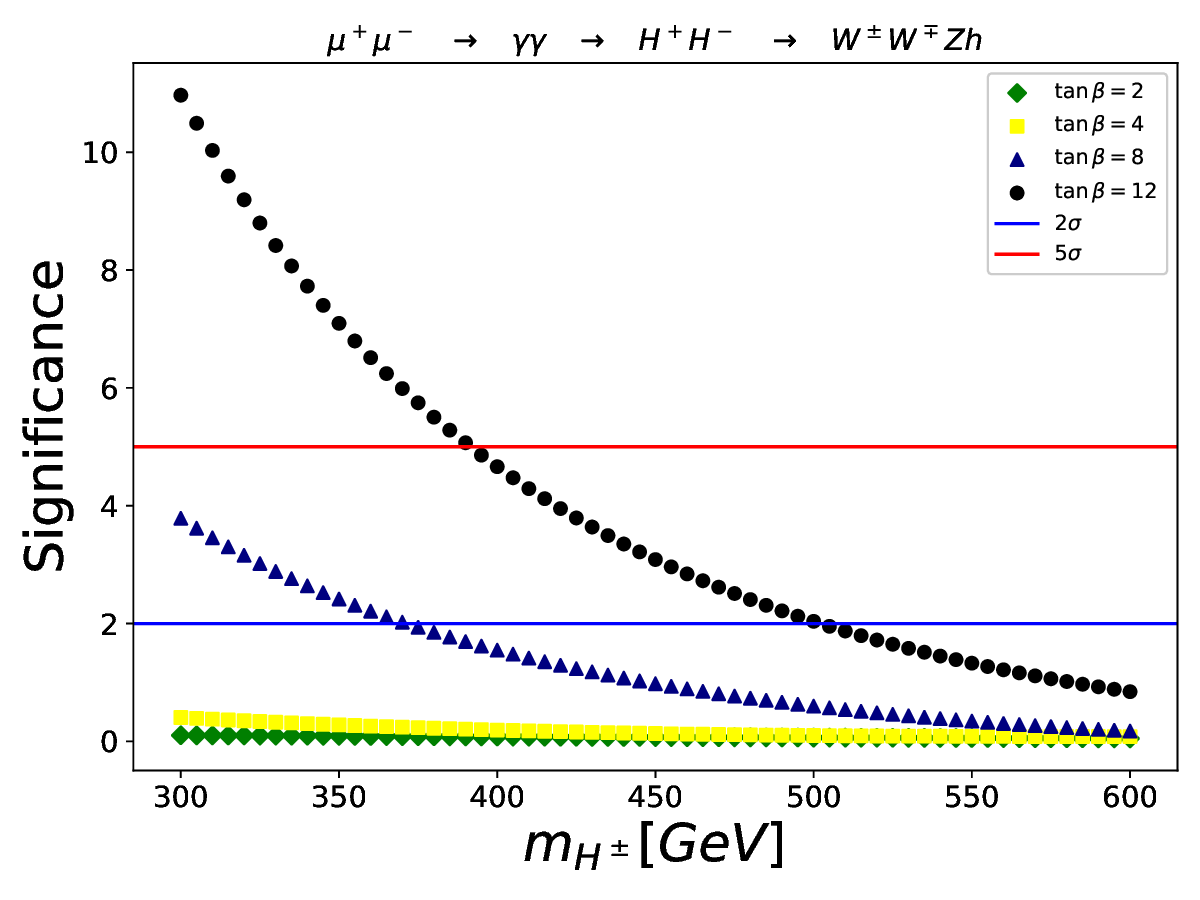}
\end{tabular}
\caption{\label{gamgamevents}
The signal events of the process
$\mu^+\mu^- \to \gamma\gamma \rightarrow
H^{+}H^{-} \rightarrow W^{\pm}W^{\mp}Zh$  
at an integrated luminosity of  
$\mathcal{L}=3000~\text{fb}^{-1}$ are
shown in the left panel. The corresponding
signal significance taking into account
the SM backgrounds is presented in
the right panel.}
\end{figure}

We now turn our attention to another process,
$\mu^+\mu^- \rightarrow \gamma \gamma \rightarrow
H^{+}H^{-} \rightarrow W^{\pm}W^{\mp}Zh$. The
total cross section is calculated by convoluting
the partonic process $\gamma \gamma \rightarrow
H^{+}H^{-} \rightarrow W^{\pm}W^{\mp}Zh$ with
the photon structure function, as follows:
\begin{equation}
\label{totalgamgam}
\sigma(s)
=
\int_{\tfrac{2m_{H^\pm}}
{\sqrt{s}}}^{x_{\text{max}}}
dz \,
\left( 2z
\int_{z^2/x_{\text{max}}}^{x_{\text{max}}}
\frac{dx}{x} \, f_{\gamma/\mu}(x)
\, f_{\gamma/\mu}
\left(z^2/x\right)
\right)
\, \hat{\sigma}(\hat{s} = z^2 s).
\end{equation}
Here, the photon structure function
$f_{\gamma/\mu}(x)$ is used, with $x$
denoting the energy fraction of the
photon emitted by the incoming lepton.
The explicit formulas for $f_{\gamma/\mu}(x)$
are given in~\cite{Zarnecki:2002qr}.
In the master formulas, we adopt
$x_{\rm max} = 0.83$ as in~\cite{Telnov:1989sd}.
The partonic process $\gamma \gamma \to H^+ H^-$
are generated by FeynArt/FormCalc~\cite{Hahn:2000kx}. 
The SM background which is the process
$\gamma \gamma \to W^\pm W^\mp Z h$,
is calculated using 
the {\tt GRACE} 
program~\cite{Belanger:2003sd}.  
To reduce the SM background, we apply cuts on the  
invariant masses of the final-state particles:  
$|m_{Wh}-m_{H^\pm}| < 10$~GeV and  
$|m_{WZ}-m_{H^\pm}| < 10$~GeV.  
In Figs.~\ref{gamgamevents}, the numbers of events 
of the process $\mu^+\mu^- \to \gamma\gamma 
\rightarrow H^{+}H^{-} \rightarrow W^{\pm}W^{\mp}Zh$ 
at an integrated luminosity of 
$\mathcal{L}=3000~\text{fb}^{-1}$ are shown in the left panel.  
The corresponding signal significance, taking into account  
the Standard Model backgrounds, is presented in the right panel.
We observe that the events become significant  
when the charged Higgs masses are in the low-mass region  
and $t_{\beta} = 8, 12$,  
whereas they are small and can be ignored in other regions.  
For charged Higgs masses in the low-mass region  
and $t_{\beta} = 12$, the significance can exceed $5\sigma$.  
In other cases, the significances are negligible.
\section{Conclusions} 
In this article, we have calculated
one-loop contributions for the
decay process $H^{\pm} \rightarrow W^{\pm} Z$
in the Two-Higgs-Doublet Model and
examined the posibility searches for charged
Higgs pair production at future
muon–TeV colliders. The computations have
performed in the $\mathcal{R}_{\xi}$ gauge,
where the analytical results are verified
through self-consistency tests such as
$\xi$-independence, ultraviolet finiteness,
and renormalization-scale stability of
the process amplitude. The numerical
results demonstrate good stability.
We have revisited
the parameter scan for the Type-X THDM
in the phenomenological results.
Based on the updated viable parameter space,
we have analyzed charged Higgs pair production
at future muon–TeV colliders by considering
the processes $\mu^+\mu^- \rightarrow H^+H^-
\rightarrow W^{\pm}W^{\mp}Zh$ and
$\mu^+\mu^- \rightarrow \gamma\gamma
\rightarrow H^+H^- \rightarrow W^{\pm}W^{\mp}Zh$.
The corresponding signal events and statistical
significances are simulated with respect to
the relevant SM backgrounds. Our findings
show that the signal significance can
exceed $5\sigma$ at several benchmark
points within the viable parameter
space of the Type-X THDM.
\\

\noindent
{\bf Acknowledgment:}~
This research is funded by Vietnam
National Foundation for Science and
Technology Development (NAFOSTED) under
the grant number $103.01$-$2023.16$.
\section*{Appendix A: One-loop form factors
in the general $\mathcal{R}_{\xi}$-gauge}
The detailed expressions for these form factors  
in the general $R_{\xi}$ gauge are given  
in the following paragraphs. In the analytical  
expressions below, we have used the notations 
as follows:
\begin{eqnarray}
A_{ij\cdots}(P) &=& A_{ij\cdots}(M_P^2),\\
B_{ij\cdots}(p^2; P_1, P_2) &=& B_{ij\cdots}(p^2; 
M_{P_1}^2,M_{P_2}^2 ),\\
C_{ij\cdots}(p_1^2, p_2^2, p_3^2; P_1, P_2,P_3) 
&=& C_{ij\cdots}(p_1^2, p_2^2, p_3^2; M_{P_1}^2, 
M_{P_2}^2, M_{P_3}^2).
\end{eqnarray}
The analytic formulas for all form factors
are expressed in terms of PV-functions,
using the adopted shorthand notations
as presented in the following paragraphs.
\subsection*{
\underline{Form factors
$\mathcal{T}^{B}_{i, \text{Trig}}$:}
}
\begin{center}
\begin{figure}[H]
\centering
\includegraphics[width=10cm, height=3cm]
{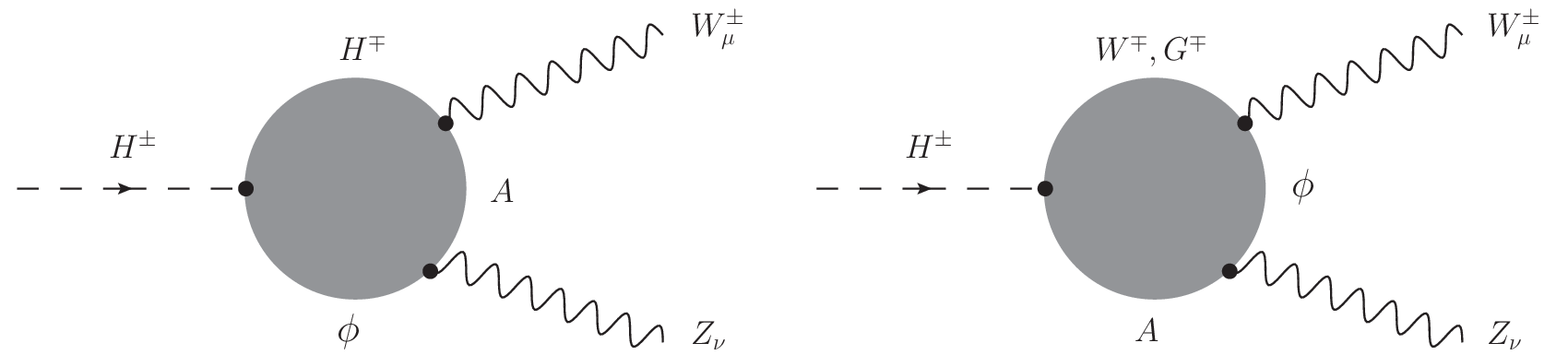}
\caption{
One-loop triangle Feynman diagrams with
$\phi = h, H, A$, together with
$H^\pm$ or $W^\pm$ propagating in the loop.
}
\label{Figs:TrigBphiA0}
\end{figure}
\end{center}

\begin{center}
\begin{figure}[H]
\centering
\includegraphics[width=10cm, height=3cm]
{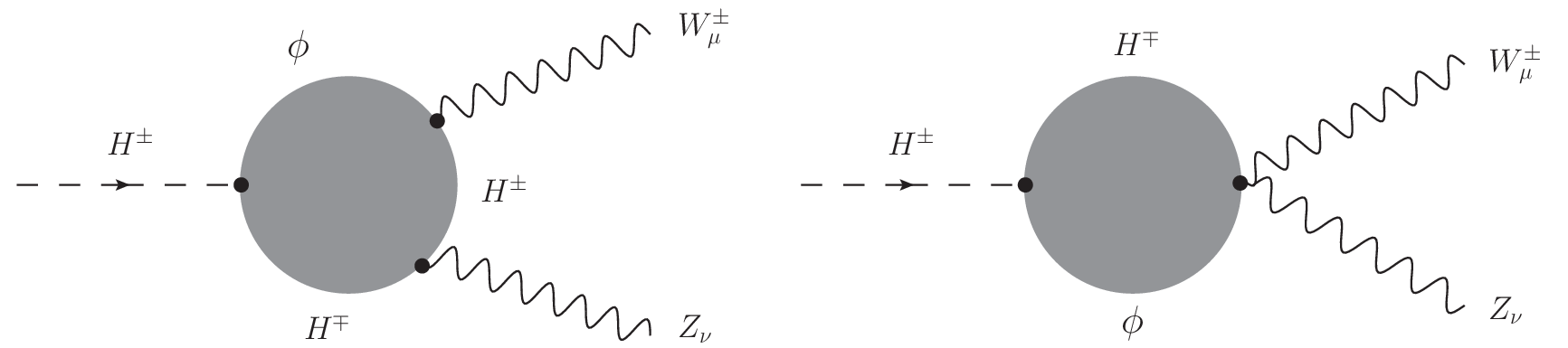}
\caption{
One-loop triangle Feynman diagrams with
$\phi = h, H$ and $H^\pm$ in the loop.
}
\label{Figs:TrigBphiHpm}
\end{figure}
\end{center}

\begin{center}
\begin{figure}[H]
\centering
\includegraphics[width=14cm, height=3cm]
{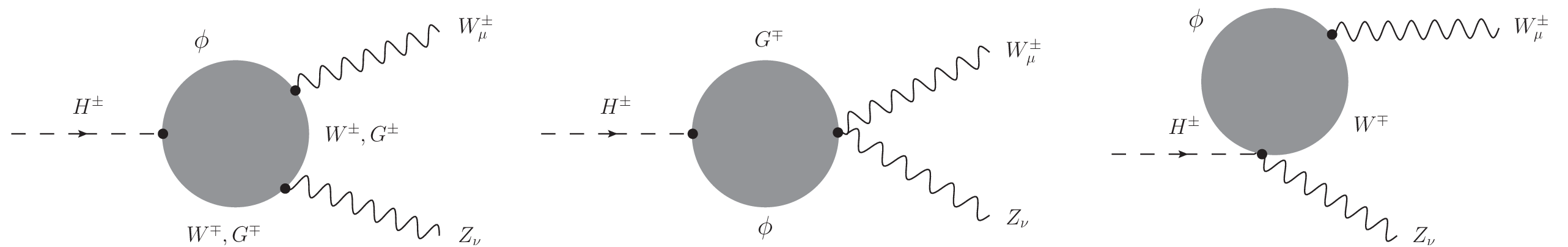}
\caption{One-loop triangle diagrams with
$\phi = h, H$ and $W^\pm$ in the loop. }
\label{Figs:TrigBphiWpm}
\end{figure}
\end{center}

\begin{center}
\begin{figure}[H]
\centering
\includegraphics[width=10cm, height=3cm]
{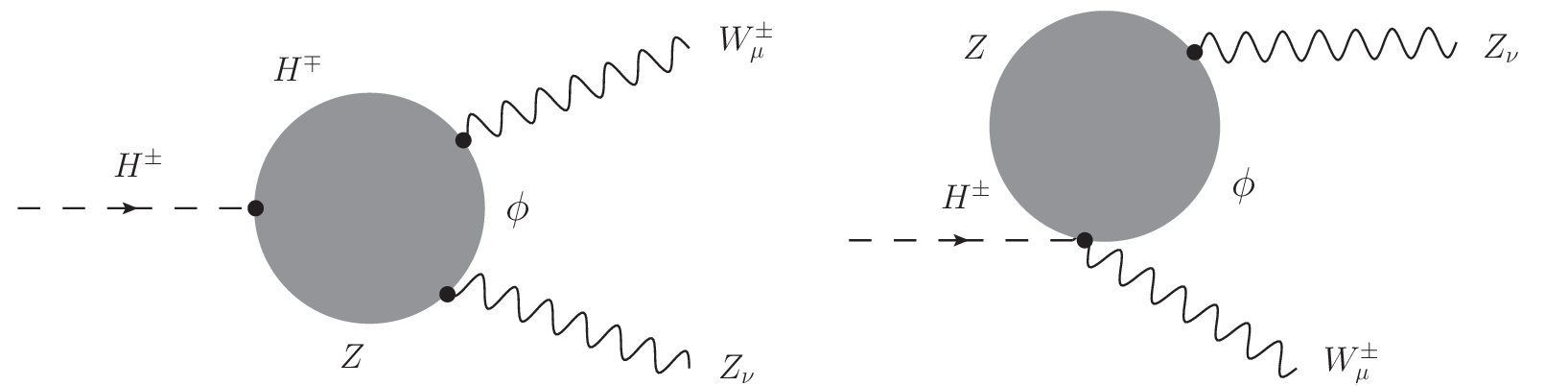}
\caption{One-loop triangle diagrams with
$\phi = h, H$ and $Z$ as internal
lines in the loop.
}
\label{Figs:TrigBphiZ}
\end{figure}
\end{center}

\begin{center}
\begin{figure}[H]
\centering
\includegraphics[width=10cm, height=3cm]
{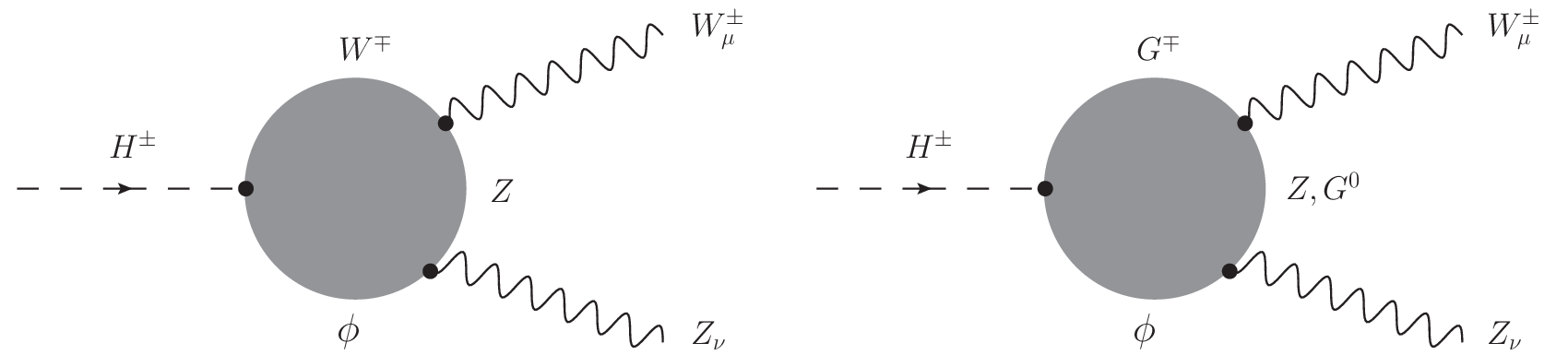}
\caption{
One-loop triangle Feynman diagrams with $\phi = h, H$
exchanging in association with $W^\pm$ and $Z$
contributions in the loop.
}
\label{Figs:TrigBphiWpmZ}
\end{figure}
\end{center}

The form factor
$\mathcal{T}^{B}_{i,\text{Trig}}$ ($i = 1,2,3$)
is expressed in terms of the main contributions
involving scalar Higgs bosons $\phi \equiv h, H$,
together with charged particles such as
$H^\pm, W^\pm, G^\pm$, and neutral particles
including $A, Z, G^0$ circulating in the loop.
\begin{eqnarray}
\label{Eq:NiBTrig}
\mathcal{T}^{B}_{i, \text{Trig}}
&=&
\sum
\limits_{\phi = h, H}
\Big(
\mathcal{T}^{B, \phi-A}_{i, \text{Trig}}
+
\mathcal{T}^{B, \phi-H^\pm}_{i, \text{Trig}}
+
\mathcal{T}^{B, \phi-W^\pm}_{i, \text{Trig}}
+
\mathcal{T}^{B, \phi-Z}_{i, \text{Trig}}
+
\mathcal{T}^{B, \phi-W^\pm Z}_{i, \text{Trig}}
\Big).
\end{eqnarray}
Each contribution of
$\mathcal{T}^{B}_{1, \text{Trig}}$
and $\mathcal{T}^{B}_{2, \text{Trig}}$
is calculated from
the corresponding Figs.~\ref{Figs:TrigBphiA0},
\ref{Figs:TrigBphiHpm}, \ref{Figs:TrigBphiWpm},
\ref{Figs:TrigBphiZ} and \ref{Figs:TrigBphiWpmZ},
respectively.
Whereas, the form factor
$\mathcal{T}^{B}_{3, \text{Trig}}$
has no contribution
in one-loop boson-exchanging
diagrams. In the
Fig.~\ref{Figs:TrigBphiA0},
the contribution in first for
exchanging of $\phi$ and
the pseudo-scalar Higgs $A$
are reading as follows:
\begin{eqnarray}
\dfrac{
\mathcal{T}^{B, \phi-A}_{1, \text{Trig}}
}{
g_{A H^- W^+}
\cdot
g_{\phi A Z}
}
&=&
-
\dfrac{
g_{\phi W^\pm W^\mp}
}{ 8\pi^2 \cdot m_W^2 }
\big[
\big(
m_{A}^2
+m_{H^\pm}^2
-\xi_{W}
m_{W}^2
\big)
C_{00}
+
2 m_{H^\pm}^2
C_{002}
\\
&&
\hspace{2cm}
+
\big(
m_{H^\pm}^2
-m_{W}^2
+m_{Z}^2
\big)
C_{001}
\big]
(Z,W,H^\pm; A,\phi,\xi_W W )
\nonumber \\
&&
+
\dfrac{
g_{\phi W^\pm W^\mp}
}{
8\pi^2
\cdot m_W^2
}
\big[
2 m_{H^\pm}^2
C_{00}
+
\big(
m_{H^\pm}^2
-m_{W}^2
+m_{Z}^2
\big)
C_{001}
\nonumber \\
&&
\hspace{4.5cm}
+
2 m_{H^\pm}^2
C_{002}
\big]
(Z,W,H^\pm;A,\phi,W)
\nonumber \\
&&
-
\dfrac{
g_{\phi H^\pm H^\mp}
}{
4\pi^2
}
C_{00}(Z,H^\pm,W;A, \phi,H^\pm),
\nonumber\\
\dfrac{
\mathcal{T}^{B, \phi-A}_{2, \text{Trig}}
}{
g_{A H^- W^+}
\cdot
g_{\phi A Z}
}
&=&
\dfrac{
g_{\phi W^\pm W^\mp}
}{8\pi^2
\cdot
m_W^2}
\Big[
2 \big(
C_{00} + C_{001}
\big)
+
\big(
m_{H^\pm}^2
-m_{W}^2
+m_{Z}^2
\big)
C_{112}
\\
&&
\hspace{0.7cm}
+
\big(
3 m_{H^\pm}^2
-m_{W}^2
+m_{Z}^2
\big)
\big(
C_{12}
+ C_{122}
\big)
+
4 \big(
m_{H^\pm}^2
C_{22}
+ C_{002}
\big)
\nonumber \\
&&
\hspace{0.7cm}
+
2 \big(
m_{H^\pm}^2
-m_{W}^2
\big)
C_{2}
+
2 m_{H^\pm}^2
C_{222}
\Big]
(Z,W,H^\pm; A,\phi,W)
\nonumber \\
&&
+
\dfrac{
g_{\phi W^\pm W^\mp}
}{8\pi^2
\cdot m_W^2}
\Big[
\big(
m_{W}^2
-m_{Z}^2
-3 m_{H^\pm}^2
\big)
C_{122}
-
2 \big(
C_{00}
+ C_{001}
+ 2 C_{002}
\big)
\nonumber\\
&&
\hspace{0.9cm}
-
\big(
m_{A}^2
+m_{H^\pm}^2
-\xi_{W}
m_{W}^2
\big)
C_{2}
-
\big(
m_{H^\pm}^2
-m_{W}^2
+m_{Z}^2
\big)
C_{112}
\nonumber\\
&&
\hspace{0.9cm}
-
\Big(
m_{A}^2
+2 m_{H^\pm}^2
-
\big(
\xi_{W}+1
\big)
m_{W}^2
+
m_{Z}^2
\Big)
C_{12}
-
2 m_{H^\pm}^2
C_{222}
\nonumber \\
&&
\hspace{0.9cm}
-
\big(
m_{A}^2
+3 m_{H^\pm}^2
-\xi_{W}
m_{W}^2
\big)
C_{22}
\Big](Z,W, H^\pm; A, \phi, \xi_W W)
\nonumber \\
&&
+
\dfrac{
g_{\phi H^\pm H^\mp}
}{4\pi^2}
C_{12}
(Z,H^\pm,W; A, \phi, H^\pm),
\nonumber \\
\dfrac{
\mathcal{T}^{B, \phi-A}_{3, \text{Trig}}
}{
g_{A H^- W^+}
\cdot
g_{\phi A Z}
}
&=&
0.
\end{eqnarray}  
Following Fig.~\ref{Figs:TrigBphiHpm},
the one-loop contributions from the
exchange of
$\phi$ and the charged Higgs $H^\pm$
are expressed as follows:
\begin{eqnarray}
\dfrac{
\mathcal{T}^{B, \phi-H^\pm}_{1, \text{Trig}}
}{
16\pi^2
\cdot
g_{\phi H^\pm H^\mp}
}
&=&
g_{\phi H^-W^+Z}
\cdot
B_{0}(H^\pm;\phi,H^\pm)
\\
&&
-
4 \,
g_{\phi H^- W^+}
\cdot
g_{Z H^\pm H^\mp}
\,
C_{00}(W,Z,H^\pm;\phi,H^\pm, H^\pm),
\nonumber \\
\dfrac{
\mathcal{T}^{B, \phi-H^\pm}_{2, \text{Trig}}
}{
16\pi^2
\cdot
g_{\phi H^\pm H^\mp}
}
&=&
-4
g_{\phi H^- W^+}
\cdot
g_{Z H^\pm H^\mp}
\big(
C_{2}
+
C_{12}
+
C_{22}
\big)
(W,Z, H^\pm; \phi,H^\pm, H^\pm),
\\
\dfrac{
\mathcal{T}^{B, \phi-H^\pm}_{3, \text{Trig}}
}{
16\pi^2
\cdot
g_{\phi H^\pm H^\mp} } &=& 0.
\end{eqnarray}
From Fig.~\ref{Figs:TrigBphiWpm},
we obtain the form factors involving
the neutral scalar Higgs boson $\phi$
in association with the vector boson
$W^\pm$ and the Goldstone bosons
$G^\pm$ in the loop. The factor
is given by
\begin{eqnarray}
\dfrac{
\mathcal{T}^{B, \phi-W^\pm}_{1, \text{Trig}}
}
{
g_{\phi W^\pm W^\mp}
}
&=&
\dfrac{
g_{\phi H^- W^+}
\cdot
g_{Z W^\pm W^\mp}
}{32\pi^2
\cdot
m_{W}^4}
\Big\{
2 m_{W}^2
\Big[
A_{0}(W)
-
A_{0}(\xi_W W)
\Big]
\\
&&
+
\Big[
2 m_{W}^2
\big(
m_{H^\pm}^2
-m_{\phi}^2
+m_{Z}^2
-m_{W}^2
\big)
B_{0}
-
2 m_{Z}^2
B_{00}
\Big]
(Z; W,\xi_{W} W)
\nonumber \\
&&
+
\Big[
2 m_{W}^2
\big(
m_{\phi}^2
-m_{H^\pm}^2
-m_{Z}^2
\big)
B_{0}
+
2 \big(
m_{Z}^2
-m_{W}^2
\big)
B_{00}
\Big](Z; W, W)
\nonumber \\
&&
+
2 m_{W}^2
B_{00}(Z;\xi_{W} W, \xi_{W} W )
-
s_{W}^2
m_{Z}^2
\big(
m_{\phi}^2
-m_{H^\pm}^2
\big)
B_{0}
(H^\pm; \phi, \xi_W W)
\nonumber \\
&&
+
2 m_{W}^2
\Big[
\big(
m_{H^\pm}^2
- m_{\phi}^2
\big)
\big(
m_{H^\pm}^2
+m_{W}^2
-m_{Z}^2
\big)
C_{2}
+
2 m_{W}^2
\big(
m_{H^\pm}^2
-m_{\phi}^2
\big)
C_{1}
\nonumber \\
&&
\hspace{1.5cm}
-
m_{\phi}^2
\big(
m_{\phi}^2
-m_{H^\pm}^2
\big)
C_{0}
\Big]
(W,Z, H^\pm;\phi,W, \xi_W W)
\nonumber \\
&&
+
2 m_{W}^2
\Big[
\big(
m_{W}^2
-m_{H^\pm}^2
-m_{\phi}^2
\big)
C_{00}
-
\big(
m_{H^\pm}^2
+m_{W}^2
-m_{Z}^2
\big)
C_{002}
\nonumber \\
&&
\hspace{5cm}
-
2 m_{H^\pm}^2
C_{001}
\Big](H^\pm, Z, W;
\phi,W, \xi_W W)
\Big\}
\nonumber \\
&&
+
\dfrac{
g_{\phi H^- W^+}
\cdot
g_{Z W^\pm W^\mp}
}{32\pi^2
\cdot
m_{W}^4}
\Big\{
2 m_{W}^2
\Big[
\big(
m_{H^\pm}^2
+m_{W}^2
-m_{Z}^2
\big)
C_{001}
+
2 m_{H^\pm}^2
C_{002}
\Big]
\nonumber \\
&&
+
2 m_{W}^2
\Big[
m_{\phi}^4
-
m_{\phi}^2
\big(
m_{H^\pm}^2
+m_{Z}^2
\big)
-
\big(
m_{H^\pm}^2
-m_{W}^2
\big)
\big(
m_{W}^2
-m_{Z}^2
\big)
\Big]
C_{0}
\nonumber \\
&&
+
2 \Big[
m_{\phi}^2
\big(
m_{Z}^2
-m_{W}^2
\big)
-
m_{Z}^2
\big(
m_{H^\pm}^2
+m_{W}^2
\big)
+
m_{W}^2
\big(
3 m_{H^\pm}^2
+m_{W}^2
\big)
\Big]
C_{00}
\nonumber \\
&&
+
2 m_{W}^4
\big(
2 m_{\phi}^2
-m_{H^\pm}^2
-m_{W}^2
-m_{Z}^2
\big)
C_{1}
\nonumber\\
&&
+
2 m_{W}^2
\Big[
m_{\phi}^2
\big(
m_{H^\pm}^2
+m_{W}^2
-m_{Z}^2
\big)
+
m_{H^\pm}^2
\big(
m_{Z}^2
-m_{H^\pm}^2
\big)
\nonumber \\
&&
\hspace{4cm}
+
m_{W}^2
\big(
m_{Z}^2
-m_{W}^2
\big)
\Big]
C_{2}
\Big\}
(W,Z,H^\pm;\phi, W, W)
\nonumber\\
&&
-
\dfrac{
g_{\phi H^- W^+ Z}
}{16\pi^2
\cdot
m_{W}^2}
\Big[
\big(
m_{W}^2
B_{0}
-
B_{00}
\big)
(W;\phi,W)
+
B_{00}
(W,\phi,\xi_W W)
\Big],
\nonumber \\
\dfrac{
\mathcal{T}^{B, \phi-W^\pm}_{2,
\text{Trig}}
}
{
g_{\phi W^\pm W^\mp}
}
&=&
-
\dfrac{
g_{\phi H^- W^+}
\cdot
g_{Z W^\pm W^\mp}
}{8\pi^2
\cdot
m_{W}^2}
\Big\{
\big(
B_{0}
+
B_{1}
\big)
(Z;W,\xi_{W} W )
+
B_{1}(Z;W, W)
\Big\}
\\
&&
-
\dfrac{
g_{\phi H^- W^+}
\cdot
g_{Z W^\pm W^\mp}
}{16\pi^2
\cdot
m_{W}^4}
\times
\nonumber\\
&&
\times
\Big\{
4 m_{W}^4
C_{1}
+
\Big[
m_{W}^2
\big(
3 m_{\phi}^2
-m_{H^\pm}^2
+m_{W}^2
\big)
-
m_{Z}^2
\big(
m_{\phi}^2
-m_{H^\pm}^2
+m_{W}^2
\big)
\Big]
C_{2}
\nonumber \\
&&
+
\Big[
m_{\phi}^2
\big(
m_{W}^2
-m_{Z}^2
\big)
+
m_{H^\pm}^2
\big(
m_{Z}^2
-2 m_{W}^2
\big)
\Big]
C_{12}
+
2 m_{W}^2
\big(
C_{00}
-
C_{001}
-
2 C_{002}
\big)
\nonumber \\
&&
+
\big(
m_{W}^2-m_{Z}^2
\big)
\big(
m_{\phi}^2
-m_{H^\pm}^2
-m_{W}^2
\big)
C_{22}
-
m_{W}^2
\big(
m_{H^\pm}^2
+m_{W}^2
-m_{Z}^2
\big)
C_{112}
\nonumber \\
&&
+
m_{W}^2
\big(
m_{Z}^2
-m_{W}^2
-3 m_{H^\pm}^2
\big)
C_{122}
-
2 m_{H^\pm}^2
m_{W}^2
C_{222}
\Big\}
(W,Z,H^\pm;\phi,W,W )
\nonumber \\
&&
-
\dfrac{
g_{\phi H^- W^+}
\cdot
g_{Z W^\pm W^\mp}
}{16\pi^2
\cdot
m_{W}^4}
\times
\nonumber\\
&&\times
\Big\{
m_{W}^2
\big(
m_{W}^2
-m_{\phi}^2
-m_{H^\pm}^2
\big)
C_{1}
+
m_{W}^2
\big(
m_{\phi}^2
-m_{H^\pm}^2
-m_{W}^2
\big)
C_{11}
\nonumber \\
&&
+
m_{W}^2
\big(
m_{\phi}^2
-2 m_{W}^2
+m_{Z}^2
\big)
C_{12}
-
2 m_{W}^2
\big(
C_{00}
-
C_{002}
-
2 C_{001}
\big)
\nonumber \\
&&
+
m_{W}^2
\big(
m_{W}^2
-m_{Z}^2
+3 m_{H^\pm}^2
\big)
C_{112}
\nonumber \\
&&
+
m_{W}^2
\big(
m_{H^\pm}^2
+m_{W}^2
-m_{Z}^2
\big)
C_{122}
+
2 m_{H^\pm}^2
m_{W}^2
C_{111}
\Big\}
(H^\pm,Z, W;
\phi, W, \xi_{W} W),
\nonumber \\
\dfrac{
\mathcal{T}^{B, \phi-W^\pm}_{3, \text{Trig}}
}{
g_{\phi W^\pm W^\mp}
}
&=&
0.
\end{eqnarray}
In Fig.~\ref{Figs:TrigBphiZ},
the corresponding one-loop form factors,
with both Higgs bosons $\phi$ and
the vector boson $Z$ exchanged in the loop,
are given as follows:
\begin{eqnarray}
\dfrac{
\mathcal{T}^{B, \phi-Z}_{1, \text{Trig}}
}
{
g_{\phi ZZ}
}
&=&
-
\dfrac{
g_{\phi H^- W^+ Z}
}{
16\pi^2
\cdot
m_{Z}^2}
\Big[
\big(
m_{Z}^2
B_{0}
-
B_{00}
\big)
(Z;\phi,Z)
+
B_{00}
(Z;\phi,\xi_Z Z)
\Big]
\\
&&
-
\dfrac{
g_{\phi H^- W^+}
\cdot
g_{Z H^\pm H^\mp}
}{
8\pi^2
\cdot
m_{Z}^2}
\times 
\nonumber\\
&&
\Big[
B_{00}
(Z; \phi,Z)
-
B_{00}
(Z;\phi,\xi_Z Z)
-
m_{Z}^2
C_{00}
(W,H^\pm,Z;\phi,H^\pm,Z)
\Big],
\nonumber\\
\dfrac{
\mathcal{T}^{B, \phi-Z}_{2, \text{Trig}}
}
{
g_{\phi ZZ}
}
&=&
-
\dfrac{
g_{\phi H^- W^+}
\cdot
g_{Z H^\pm H^\mp}
}{16\pi^2}
\big(
4 C_{2} + 2 C_{12}
\big)
(W,H^\pm,Z;\phi,H^\pm,Z),
\\
\dfrac{
\mathcal{T}^{B, \phi-Z}_{3, \text{Trig}}
}
{
g_{\phi ZZ}
}
&=&
0.
\end{eqnarray}
In Fig.~\ref{Figs:TrigBphiWpmZ},
the form factors corresponding to the
neutral scalar Higgs $\phi$, accompanied by
both the vector bosons $W^\pm$ and $Z$
and the Goldstone bosons $G^\pm$ and $G^0$
in the loop, are represented as follows:
\begin{eqnarray}
\dfrac{
\mathcal{T}^{B, \phi-W^\pm Z}_{1, \text{Trig}}
}
{
g_{\phi ZZ}
}
&=&
\dfrac{
g_{\phi H^- W^+}
\cdot
g_{Z W^\pm W^\mp}
}{16\pi^2}
\xi_{W}
B_{0}
(Z; \phi,Z)
\\
&&
+
\dfrac{
g_{\phi H^- W^+}
\cdot
g_{Z W^\pm W^\mp}
}{16\pi^2
\cdot
m_{W}^2 m_{Z}^2}
\Big\{
\big[
c_{W}^2
m_{Z}^2
\big(
m_{\phi}^2
-m_{H^\pm}^2
\big)
+
m_{W}^2
\big(
2 m_{H^\pm}^2
-
\xi_{W}
m_{W}^2
\big)
\big]
C_{00}
\nonumber \\
&&
+
m_{W}^2
\big[
\big(
m_{H^\pm}^2
-m_{W}^2
+m_{Z}^2
\big)
C_{002}
+
2 m_{H^\pm}^2
C_{001}
\big]
\Big\}
(H^\pm, W, Z;
\phi,\xi_{W} W, \xi_{Z} Z )
\nonumber \\
&&
+
\dfrac{
g_{\phi H^- W^+}
\cdot
g_{Z W^\pm W^\mp}
}{16\pi^2
\cdot
m_{W}^2 m_{Z}^2}
\Bigg\{
m_{Z}^2
\Big[
m_{H^\pm}^2
\big[
m_{W}^2
\big(
2-3 \xi_{W}
\big)
+
m_{Z}^2
\big(
s_{W}^2-2
\big)
+
2 m_{H^\pm}^2
\big]
\nonumber\\
&&
\hspace{4.5cm}
-
m_{\phi}^2
m_{Z}^2
s_{W}^2
+
\xi_{W}
m_{W}^2
\big[
m_{W}^2
\big(
\xi_{W}-1
\big)
+
m_{Z}^2
\big]
\Big]
C_{0}
\nonumber\\
&&
-
m_{Z}^2
\Big[
m_{H^\pm}^2
\big(
2 m_{W}^2
+ 2 m_{Z}^2-
3 m_{H^\pm}^2
\big)
\nonumber \\
&&
\hspace{4.5cm}
+
2 \xi_{W}
m_{W}^2
\big(
m_{H^\pm}^2
-
m_{W}^2
\big)
+
\big(
m_{W}^2
-m_{Z}^2
\big)^2
\Big]
C_{1}
\nonumber\\
&&
+
\Big[
m_{Z}^2
\big(
m_{\phi}^2
s_{W}^2
+2 m_{W}^2
-2 m_{Z}^2
\big)
\nonumber \\
&&
\hspace{2.5cm}
-
m_{H^\pm}^2
\big[
2 m_{W}^2
+
m_{Z}^2
\big(
s_{W}^2-4
\big)
\big]
+
\xi_{W}
m_{W}^2
\big(
m_{W}^2
-m_{Z}^2
\big)
\Big]
C_{00}
\nonumber \\
&&
-
m_{Z}^2
\Big[
m_{H^\pm}^2
\big(
2 m_{W}^2
+ 2 m_{Z}^2
-3 m_{H^\pm}^2
\big)
+
\big(
m_{W}^2
-m_{Z}^2
\big)^2
\Big]
C_{12}
\nonumber \\
&&
+
m_{Z}^2
\Big[
\xi_{W}
m_{W}^2
\big(
m_{Z}^2
-m_{W}^2
-3 m_{H^\pm}^2
\big)
+
4 m_{H^\pm}^2
\big(
m_{H^\pm}^2
+m_{W}^2
-m_{Z}^2
\big)
\Big]
C_{2}
\nonumber \\
&&
+
\big(
m_{Z}^2
-m_{W}^2
\big)
\Big[
2 m_{H^\pm}^2
C_{002}
+
\big(
m_{H^\pm}^2
-m_{W}^2
+m_{Z}^2
\big)
C_{001}
\Big]
\nonumber \\
&&
+
m_{Z}^2
\Big[
\big(
m_{H^\pm}^2
-m_{W}^2
+m_{Z}^2
\big)
\big(
m_{H^\pm}^2
-m_{W}^2
-m_{Z}^2
\big)
C_{11}
\nonumber \\
&&
\hspace{2.5cm}
+
2 m_{H^\pm}^2
\big(
m_{H^\pm}^2
+m_{W}^2
-m_{Z}^2
\big)
C_{22}
\Big]
\Bigg\}
(Z,W,H^\pm; \phi,Z, \xi_W W)
\nonumber \\
&&
+
\dfrac{
g_{\phi H^- W^+}
\cdot
g_{Z W^\pm W^\mp}
}{16\pi^2
\cdot
m_{W}^2}
\Bigg\{
2 \big(
m_{H^\pm}^2
+m_{W}^2
-m_{Z}^2
\big)
\Big[
\big(
m_{W}^2
- m_{H^\pm}^2
\big)
C_{0}
-
C_{00}
\Big]
\nonumber \\
&&
+
\Big[
m_{H^\pm}^2
\big(
2 m_{W}^2
+ 2 m_{Z}^2
-3 m_{H^\pm}^2
\big)
+
\big(
m_{W}^2
-m_{Z}^2
\big)^2
\Big]
\big(
C_{2}
+
C_{12}
\big)
\nonumber \\
&&
+
\big(
m_{H^\pm}^2
-m_{W}^2
+m_{Z}^2
\big)
\Big[
\big(
m_{W}^2
+m_{Z}^2
-m_{H^\pm}^2
\big)
C_{22}
-
C_{002}
\Big]
\nonumber \\
&&
-
2 m_{H^\pm}^2
\Big[
\big(
m_{H^\pm}^2
+m_{W}^2
-m_{Z}^2
\big)
\big(
2 C_{1}
+
C_{11}
\big)
\nonumber \\
&&
\hspace{5.5cm}
+
\big(
C_{00}
+
C_{001}
\big)
\Big]
\Bigg\}
(H^\pm, W, Z; \phi, W, Z),
\nonumber \\
\dfrac{
\mathcal{T}^{B,
\phi-W^\pm Z}_{2, \text{Trig}}
}{
g_{\phi ZZ}
}
&=&
\dfrac{
g_{\phi H^- W^+}
\cdot
g_{Z W^\pm W^\mp}
}{16\pi^2
\cdot
m_{W}^2 m_{Z}^2}
\Bigg\{
\Big[
m_{Z}^2 c_{W}^2
\big(
m_{\phi}^2
-
m_{H^\pm}^2
\big)
+
m_{W}^2
\big(
2 m_{H^\pm}^2
-
\xi_{W}
m_{W}^2
\big)
\Big]
C_{1}
\nonumber \\
&&
+
\Big[
m_{Z}^2 c_{W}^2
\big(
m_{\phi}^2
-
m_{H^\pm}^2
\big)
+
m_{W}^2
\big(
4 m_{H^\pm}^2
-
\xi_{W}
m_{W}^2
\big)
\Big]
C_{11}
\nonumber\\
&&
+
\Big[
m_{Z}^2
\big(
c_{W}^2
m_{\phi}^2
+
m_{W}^2
\big)
-
m_{H^\pm}^2
\big(
c_{W}^2
m_{Z}^2
-
3 m_{W}^2
\big)
-
m_{W}^4
\big(
\xi_{W} + 1
\big)
\Big]
C_{12}
\nonumber \\
&&
+
2m_{W}^2 m_{H^\pm}^2
C_{111}
+
m_{W}^2
\big(
3 m_{H^\pm}^2
-m_{W}^2
+m_{Z}^2
\big)
C_{112}
\nonumber\\
&&
+
m_{W}^2
\big(
m_{H^\pm}^2
-m_{W}^2
+m_{Z}^2
\big)
C_{122}
\nonumber \\
&&
+
2 m_{W}^2
\big(
C_{00}
+
2 C_{001}
+
C_{002}
\big)
\Bigg\}
(H^\pm, W, Z;
\phi,\xi_{W} W,\xi_{Z} Z )
\nonumber \\
&&
+
\dfrac{
g_{\phi H^- W^+}
\cdot
g_{Z W^\pm W^\mp}
}{16\pi^2
\cdot
m_{W}^2 m_{Z}^2}
\Bigg\{
\big(
m_{Z}^2
-m_{W}^2
\big)
\Big[
2 C_{00}
+
2 C_{001}
+
4 C_{002}
+
2 m_{H^\pm}^2
C_{222}
\nonumber \\
&&
+
\big(
m_{H^\pm}^2
-m_{W}^2
+m_{Z}^2
\big)
C_{112}
+
\big(
3 m_{H^\pm}^2
-m_{W}^2
+m_{Z}^2
\big)
C_{122}
\Big]
\nonumber \\
&&
+
\Big[
m_{\phi}^2
m_{Z}^2
s_{W}^2
-
m_{H^\pm}^2
\big(
m_{W}^2
- m_{Z}^2
\big)
+
\xi_{W}
m_{W}^2
\big(
m_{W}^2
-m_{Z}^2
\big)
\Big]
C_{2}
\nonumber \\
&&
+
\Big[
m_{\phi}^2
m_{Z}^2
s_{W}^2
-
3 m_{H^\pm}^2
\big(
m_{W}^2
- m_{Z}^2
\big)
+
\xi_{W}
m_{W}^2
\big(
m_{W}^2
-m_{Z}^2
\big)
\Big]
C_{22}
\nonumber \\
&&
-
\Big[
3 m_{H^\pm}^2
m_{W}^2
-
m_{Z}^2
\big(
m_{\phi}^2
s_{W}^2
+
m_{Z}^2
\big)
+
m_{H^\pm}^2
m_{Z}^2
\big(
s_{W}^2-3
\big)
\nonumber \\
&&
-
m_{W}^4
\big(
\xi_{W}+1
\big)
+
m_{W}^2
m_{Z}^2
\big(
\xi_{W}+2
\big)
\Big]
C_{12}
\Bigg\}
(Z,W,H^\pm; \phi,Z,\xi_W W)
\nonumber \\
&&
+
\dfrac{
g_{\phi H^- W^+}
\cdot
g_{Z W^\pm W^\mp}
}{16\pi^2
\cdot
m_{W}^2 }
\Bigg\{
\big(
m_{W}^2
-m_{Z}^2
-3 m_{H^\pm}^2
\big)
\big(
C_{12}
+
C_{112}
\big)
\nonumber\\
&&
-
\big(
m_{H^\pm}^2
-m_{W}^2
+m_{Z}^2
\big)
C_{122}
+
2 m_{W}^2
\big(
C_{1}
+
2 C_{2}
\big)
\nonumber\\
&&
-
2 m_{H^\pm}^2
\big(
C_{1}
+
2 C_{11}
+
C_{111}
\big)
-
2 \big(
C_{00}
+
2 C_{001}
+
C_{002}
\big)
\Bigg\}
(H^\pm,W,Z; \phi,W, Z),
\nonumber \\
\dfrac{
\mathcal{T}^{B,
\phi-W^\pm Z}_{3, \text{Trig}}
}
{
g_{\phi ZZ}
}
&=&
0.
\end{eqnarray}
\subsection*{
\underline{Form factors
$\mathcal{T}^{B}_{i, \text{Self}}$}:
}
Following Fig.~\ref{Figs:SelfB}, the form factors
$\mathcal{T}^{B}_{i, \text{Self}}$ for $i = 1, 2, 3$
are decomposed into two contributions from one-point
$(1P)$ and two-point $(2P)$ Feynman diagrams.
These factors are expressed as follows:
\begin{eqnarray}
\mathcal{T}^{B}_{i, \text{Self}}
&=&
\mathcal{T}^{B}_{i, \text{Self} - 1P}
+
\sum
\limits_{\phi = h, H}
\Big(
\mathcal{T}^{B, \phi-H^\pm}_{i, \text{Self} - 2P}
+
\mathcal{T}^{B, \phi-W^\pm}_{i, \text{Self} - 2P}
\Big)
\end{eqnarray}
in which, these above
form factors for $i = 1$
are reading as follows
\begin{eqnarray}
\dfrac{
\mathcal{T}^{B}_{1, \text{Self} - 1P}
}{
g_{Z W^\pm W^\mp}
}
&=&
\dfrac{m_{W} s_{W}^2}{
32\pi^2
\cdot
c_{W}^2
\big(m_{H^\pm}^2
- \xi m_{W}^2\big)
}
\times
\\
&&
\times
\Bigg\{
g_{G^0 G^0 H^- G^+}
\,
A_0
(\xi_{Z} Z)
+
2 g_{G^+ G^- H^- G^+}
\,
A_0
(\xi_{W} W)
\nonumber\\
&&
\hspace{0.5cm}
+
g_{A A H^- G^+}
\,
A_0
(A)
+
2 g_{H^+ H^- H^- G^+}
\,
A_0
(H^\pm)
\nonumber \\
&&
\hspace{1.5cm}
+
g_{h h \, H^- G^+}
\,
A_0
(h)
+
g_{H H \, H^- G^+}
\,
A_0
(H)
\Bigg\},
\nonumber \\
\dfrac{
\mathcal{T}^{B, \phi-H^\pm}_{1,
\text{Self} - 2P}
}{
g_{Z W^\pm W^\mp}
}
&=&
-
\dfrac{
g_{\phi H^- W^+}
\cdot
g_{\phi H^\pm H^\mp}
}{
16\pi^2
\cdot
m_{W}^2
\big(
m_{H^\pm}^2
- \xi_{W} m_{W}^2
\big)}
\Bigg\{
2 \xi_{W} m_{W}^2
\big(
m_{W}^2
-m_{Z}^2
\big)
B_{1}(H^\pm; \phi, H^\pm)
\\
&&
\hspace{1cm}
+
\Big[
m_{Z}^2 s_{W}^2
\big(
m_{H^\pm}^2
-
m_{\phi}^2
\big)
+
\xi_{W} m_{W}^2
\big(
m_{W}^2
-m_{Z}^2
\big)
\Big]
B_{0}(H^\pm; \phi, H^\pm)
\Bigg\},
\nonumber \\
\dfrac{\mathcal{T}^{B,
\phi-W^\pm}_{1, \text{Self} - 2P}
}{
g_{Z W^\pm W^\mp}
}
&=&
-
\dfrac{
g_{\phi H^- W^+}
\cdot
g_{\phi W^\pm W^\mp}
}{
32\pi^2
\cdot
m_{W}^4
\big(
m_{H^\pm}^2
- \xi_{W} m_{W}^2
\big)}
\times
\\
&&
\times
\Bigg\{
s_{W}^2
m_{Z}^2
\Big[
\big(
m_{\phi}^2
-m_{H^\pm}^2
\big)
A_{0}
(W)
-
\big(
m_{\phi}^2
-m_{H^\pm}^2
+\xi_{W}
m_{W}^2
\big)
A_{0}
(\xi_{W} W)
\Big]
\nonumber\\
&&
+
\big(
m_{\phi}^2
-m_{H^\pm}^2
\big)
\Big[
s_{W}^2
m_{H^\pm}^2
m_{Z}^2
+
\xi_{W}
m_{W}^2
\big(
m_{W}^2
-m_{Z}^2
\big)
\Big]
B_{0}
(H^\pm; \phi,
\xi_{W} W)
\nonumber \\
&&
+
2 m_{W}^2
\Big[
\xi_{W}
\big(
m_{W}^2
-m_{Z}^2
\big)
\big(
m_{H^\pm}^2
+m_{W}^2
-m_{\phi}^2
\big)
\nonumber\\
&&
\hspace{7cm}
+
s_{W}^2
m_{H^\pm}^2
m_{Z}^2
\Big]
B_{1}
(H^\pm; \phi,W)
\nonumber \\
&&
+
s_{W}^2
m_{Z}^2
\Big[
\big(
m_{\phi}^2
-m_{H^\pm}^2
\big)^2
-
m_{W}^2
\big(
m_{\phi}^2
+m_{H^\pm}^2
\big)
\Big]
B_{0}
(H^\pm; \phi,W)
\nonumber \\
&&
\hspace{1.5cm}
-
2 \xi_{W}
m_{W}^2
\big(
m_{W}^2
-m_{Z}^2
\big)
\big(
m_{\phi}^2
-m_{H^\pm}^2
+m_{W}^2
\big)
B_{0}
(H^\pm; \phi,W)
\Bigg\}.
\nonumber
\end{eqnarray}
Other ones for $i = 2, 3$ have
no contribution, or
\begin{eqnarray}
\label{Eq:NiBSelf}
&&
\mathcal{T}^{B}_{i, \text{Self} - 1P}
=
\mathcal{T}^{B, \phi-H^\pm}_{i, \text{Self} - 2P}
=
\mathcal{T}^{B, \phi-W^\pm}_{i, \text{Self} - 2P}
=
0.
\end{eqnarray}
\subsection*{
\underline{Form factors
$\mathcal{T}^{B}_{i, \text{Tad}}$}:
}
\begin{center}
\begin{figure}[H]
\centering
\includegraphics[width=14cm, height=6cm]
{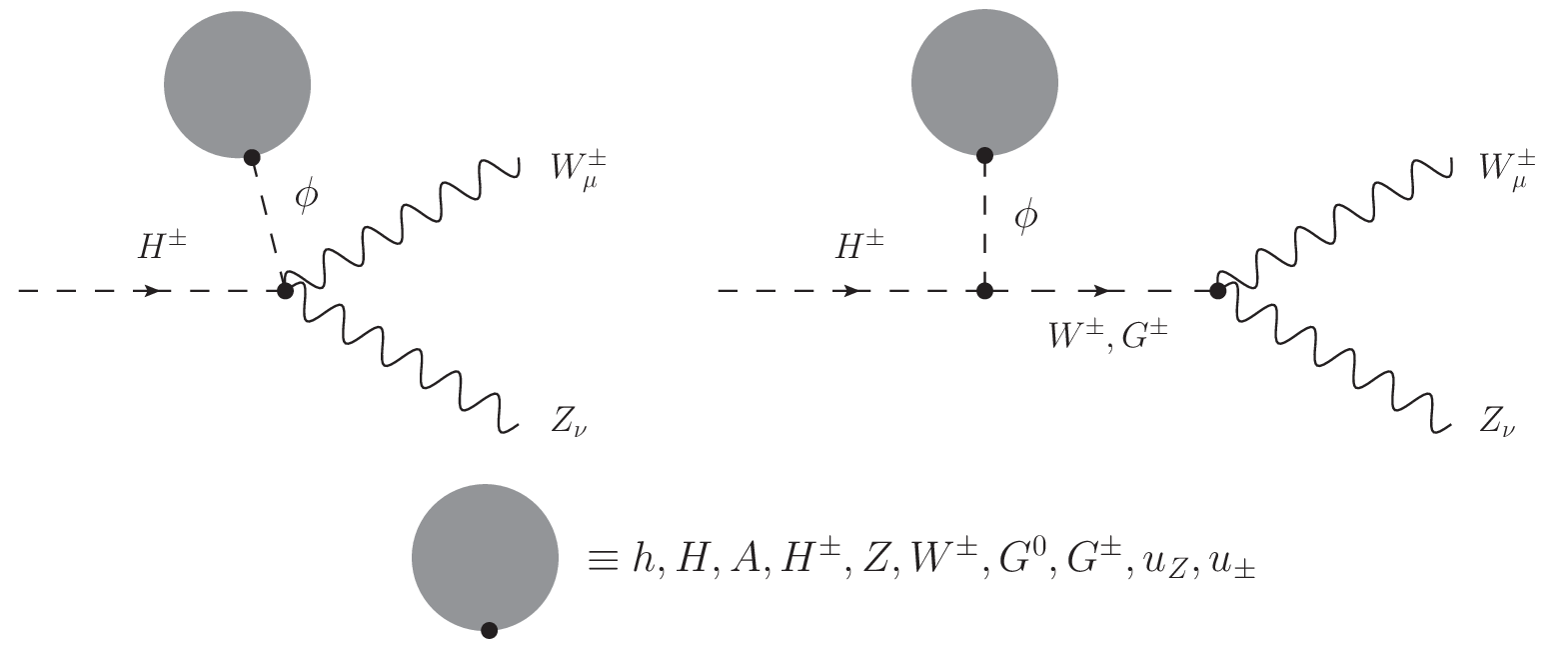}
\caption{
Tadpole Feynman diagram contributions
with poles $\phi \equiv h, H$.
}
\label{Figs:TadpoleB}
\end{figure}
\end{center}

According to Fig.~\ref{Figs:TadpoleB},
the form factor
$\mathcal{T}^{B}_{i, \text{Tad}}$ for $i = 1, 2, 3$
is expressed in terms of the scalar Higgs
$\phi \equiv h, H$ pole coupling with the
bubble diagrams. In these loops, the neutral and
pseudo-scalar Higgs particles $h, H, A$, the
charged Higgs $H^\pm$, the vector bosons $W^\pm$
and $Z$, the Goldstone bosons $G^\pm$ and $G^0$,
and the corresponding ghost particles $u_\pm$ and $u_Z$
are all taken into account. As a result, the factor
is given by
\begin{eqnarray}
\mathcal{T}^{B}_{i, \text{Tad}}
&=&
\sum
\limits_{\phi = h, H}
\mathcal{T}^{B, \phi}_{i, \text{Tad}}.
\end{eqnarray}
Where the form factors expressed
as follows:
\begin{eqnarray}
\mathcal{T}^{B, \phi}_{1, \text{Tad}}
&=&
-
\dfrac{1}{
64\pi^2
\cdot
m_{\phi}^2 
\big(m_{H^\pm}^2
- \xi_{W} m_{W}^2 \big)
}
\times
\nonumber\\
&&
\times
\Big[
g_{\phi H^- W^+ Z}
\,
\big(
m_{H^\pm}^2
- \xi_{W} m_{W}^2
\big)
\\
&&
+
\big(
m_{\phi}^2
-m_{H^\pm}^2
\big)
t_{W}^2\;
g_{\phi H^- W^+}
\cdot
g_{Z W^\pm W^\mp}
\nonumber\\
&&
+
\xi_{W} 
\big(
m_{Z}^2
-m_{W}^2
\big)
g_{\phi H^- W^+}
\cdot
g_{Z W^\pm W^\mp}
\Big]
\times
\nonumber \\
&&\times
\Bigg\{
4
m_{Z}^2
\,
g_{\phi Z Z}
+
8 m_{W}^2
\,
g_{\phi W^\pm W^\mp}
\nonumber\\
&&
+
2 g_{\phi A A}
\;
A_{0}
(A)
+
2
g_{\phi h h}
\;
A_{0}
(h)
\nonumber\\
&&
\hspace{0cm}
+
2
\; 
g_{\phi H H}
\,
A_{0}
(H)
+
4
g_{\phi H^\pm H^\mp}
\,
A_{0}
(H^\pm)
\nonumber \\
&&
\hspace{0cm}
-
12 g_{\phi W^\pm W^\mp}
A_{0}
(W)
-
6 g_{\phi Z Z}
\,
A_{0}
(Z)
\nonumber\\
&&
-
g_{\phi Z Z}
\,
\dfrac{m_{\phi}^2}{m_{Z}^2}
\;
A_{0}
(\xi_{Z} Z)
\nonumber\\
&&
-
2 
\dfrac{m_{\phi}^2}{m_{W}^2}
\,
g_{\phi W^\pm W^\mp}
\,
A_{0}
(\xi_{W} W)
\Bigg\},
\nonumber\\
\mathcal{T}^{B, \phi}_{2, \text{Tad}}
&=&
\mathcal{T}^{B, \phi}_{3, \text{Tad}} = 0.
\end{eqnarray}
\subsection*{
\underline{Form factors
$\mathcal{T}^{F}_{i, \text{Trig}}$}:
}
\begin{center}
\begin{figure}[H]
\centering
\includegraphics[width=12cm, height=4cm]
{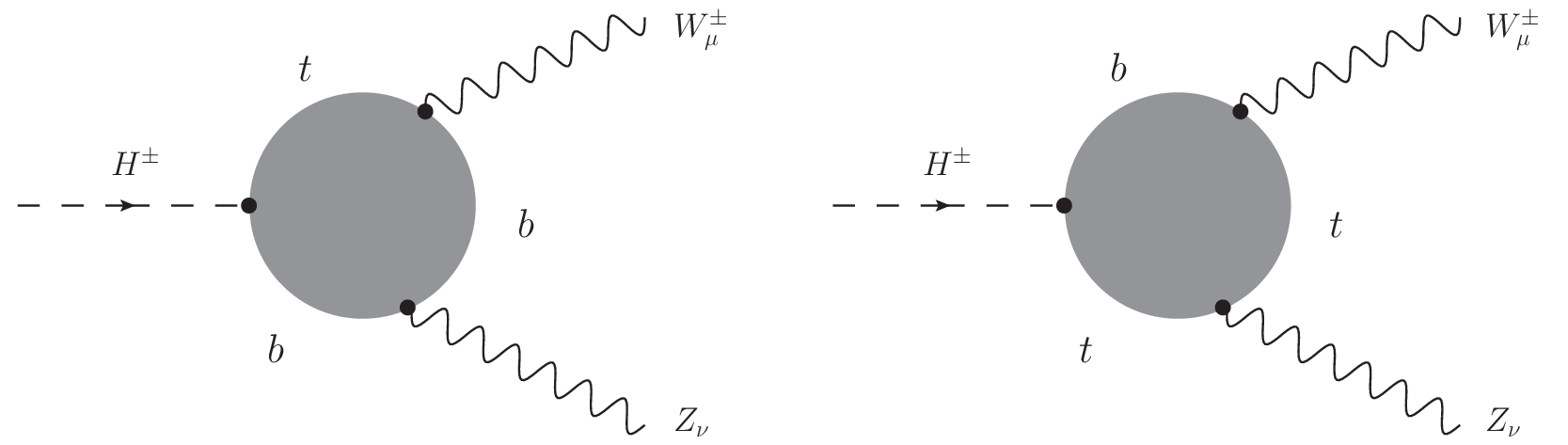}
\caption{
One-loop triangle Feynman diagrams with fermion loops.
}
\label{Figs:TrigF}
\end{figure}
\end{center}

The form factors
$\mathcal{T}^{F}_{i, \text{Trig}}$,
for $i = 1, 2, 3$ as depicted in
the corresponding
Fig.~\ref{Figs:TrigF} are given by:
\begin{eqnarray}
\dfrac{\mathcal{T}^{F}_{1,
\text{Trig}}
}{
g_{W^+ \bar{t} \, b}
}
&=&
\dfrac{
N^C_Q
}{
8\pi^2}
\cdot
\Bigg\{
g^{L}_{H^- t \bar{b}}
\,
m_{b}^2
\,
\Big[
g^{L}_{Z t \bar{t}}
\,
B_{0}
(Z; t,t)
-
\big(
g^{R}_{Z b \bar{b}}
-
g^{L}_{Z b \bar{b}}
\big)
B_{0}
(Z; b,b)
\Big]
\\
&&
-
g^{R}_{H^- t \bar{b}}
\,
m_{t}^2
\,
\Big[
\big(
g^{R}_{Z t \bar{t}}
-
g^{L}_{Z t \bar{t}}
\big)
B_{0}
(Z; t,t)
-
g^{L}_{Z b \bar{b}}
\,
B_{0}
(Z; b,b)
\Big]
\nonumber \\
&&
-
m_{t}^2
\,
m_{b}^2
\,
\Big[
g^{R}_{Z t \bar{t}}
\cdot
g^{L}_{H^- t \bar{b}}
\,
C_{0}
(W,Z,H^\pm;
b,t,t)
\nonumber\\
&&
\hspace{4.2cm}
+
g^{R}_{Z b \bar{b}}
\cdot
g^{R}_{H^- t \bar{b}}
\,
C_{0}
(Z,H^\pm,W;
b,b,t)
\Big]
\Bigg\}
\nonumber \\
&&
+
\dfrac{
N^C_Q
}{
16\pi^2}
\cdot
\Bigg\{
-
g^{L}_{Z b \bar{b}}
\cdot
g^{L}_{H^- t \bar{b}}
\,
m_{b}^2
\,
\Big[
2 \big(
m_{W}^2
- m_{t}^2
\big)
C_{0}
+
4 C_{00}
\nonumber \\
&&
\hspace{1.5cm}
+
2 m_{W}^2
C_{2}
-
\big(
m_{H^\pm}^2
-m_{W}^2
-m_{Z}^2
\big)
C_{1}
\Big]
(Z,H^\pm,W;
b,b,t)
\nonumber \\
&&
+
g^{L}_{Z t \bar{t}}
\cdot
g^{L}_{H^- t \bar{b}}
\,
m_{b}^2
\,
\Big[
\big(
2 m_{b}^2
+m_{H^\pm}^2
+m_{W}^2
-m_{Z}^2
\big)
C_{0}
\nonumber \\
&&
\hspace{3cm}
-
4 C_{00}
+
\big(
m_{H^\pm}^2
+3 m_{W}^2
-m_{Z}^2
\big)
C_{1}
\nonumber\\
&&
\hspace{3cm}
+
\big(
3 m_{H^\pm}^2
+m_{W}^2
-m_{Z}^2
\big)
C_{2}
\Big]
(W,Z,H^\pm;
b,t,t)
\nonumber \\
&&
-
g^{R}_{Z b \bar{b}}
\cdot
g^{L}_{H^- t \bar{b}}
\,
m_{b}^2
\,
\Big[
\big(
2 m_{t}^2
-m_{H^\pm}^2
-m_{W}^2
+m_{Z}^2
\big)
C_{0}
\nonumber \\
&&
\hspace{3cm}
+
\big(
m_{H^\pm}^2
-m_{W}^2
+m_{Z}^2
\big)
C_{1}
\nonumber\\
&&
\hspace{3cm}
-
\big(
m_{H^\pm}^2
+m_{W}^2
-m_{Z}^2
\big)
C_{2}
\Big]
(Z,H^\pm,W;
b,b,t)
\nonumber \\
&&
-
g^{L}_{Z b \bar{b}}
\cdot
g^{R}_{H^- t \bar{b}}
\,
m_{t}^2
\,
\Big[
2 \big(
m_{W}^2
- m_{t}^2
\big)
C_{0}
+
4 C_{00}
-
2 \big(
m_{H^\pm}^2
-m_{W}^2
\big)
C_{1}
\nonumber \\
&&
\hspace{3cm}
+
\big(
m_{H^\pm}^2
+3 m_{W}^2
-m_{Z}^2
\big)
C_{2}
\Big]
(Z,H^\pm,W;
b,b,t)
\nonumber \\
&&
+
g^{L}_{Z t \bar{t}}
\cdot
g^{R}_{H^- t \bar{b}}
\,
m_{t}^2
\,
\Big[
2 m_{b}^2
\,
C_{0}
-
4
C_{00}
+
\big(
m_{H^\pm}^2
+m_{W}^2
-m_{Z}^2
\big)
C_{2}
\nonumber \\
&&
\hspace{6cm}
+
2 m_{W}^2
C_{1}
\Big]
(W,Z,H^\pm;
b,t,t)
\nonumber \\
&&
-
g^{R}_{Z t \bar{t}}
\cdot
g^{R}_{H^- t \bar{b}}
\,
m_{t}^2
\,
\Big[
2 m_{b}^2
\,
C_{0}
+
\big(
m_{H^\pm}^2
+m_{W}^2
-m_{Z}^2
\big)
C_{1}
\nonumber \\
&&
\hspace{6cm}
+
2 m_{H^\pm}^2
C_{2}
\Big]
(W,Z,H^\pm;
b,t,t)
\Bigg\},
\nonumber \\
\dfrac{\mathcal{T}^{F}_{2, \text{Trig}}
}{
g_{W^+ \bar{t} \, b}
}
&=&
\dfrac{
N^C_Q
}{
8\pi^2}
\cdot
\Bigg\{
-
g^{L}_{Z t \bar{t}}
\cdot
g^{L}_{H^- t \bar{b}}
\,
m_{b}^2
\,
\times
\\
&&
\hspace{2cm}
\times
\big(
C_{0}
+
C_{1}
+
3 C_{2}
+
2 C_{12}
+
2 C_{22}
\big)
(W,Z,H^\pm;
b,t,t)
\nonumber \\
&&
+
g^{L}_{H^- t \bar{b}}
\,
m_{b}^2
\times
\nonumber\\
&&
\hspace{0.2cm}
\times
\Big[
g^{L}_{Z b \bar{b}}
\,
\big(
C_{1}
+
2 C_{12}
\big)
-
g^{R}_{Z b \bar{b}}
\,
\big(
C_{0}
+
C_{1}
+
C_{2}
\big)
\Big]
(Z,H^\pm,W;
b,b,t)
\nonumber \\
&&
+
g^{L}_{Z b \bar{b}}
\cdot
g^{R}_{H^- t \bar{b}}
\,
m_{t}^2
\,
\big(
C_{2}
+
2 C_{12}
\big)
(Z,H^\pm,W;
b,b,t)
\nonumber \\
&&
+
g^{R}_{H^- t \bar{b}}
\,
m_{t}^2
\,
\Big[
g^{R}_{Z t \bar{t}}
\,
C_{1}
-
g^{L}_{Z t \bar{t}}
\,
\big(
C_{2}
+
2 C_{12}
+
2 C_{22}
\big)
\Big]
(W,Z,H^\pm;
b,t,t)
\Bigg\},
\nonumber \\
\dfrac{\mathcal{T}^{F}_{3, \text{Trig}}
}{
g_{W^+ \bar{t} \, b}
}
&=&
\dfrac{
N^C_Q
}{
8\pi^2}
\times
\Bigg\{
-
g^{L}_{Z t \bar{t}}
\cdot
g^{L}_{H^- t \bar{b}}
\,
m_{b}^2
\,
\big(
C_{0}
+
C_{1}
+
C_{2}
\big)
(W,Z,H^\pm;
b,t,t)
\\
&&
+
g^{L}_{H^- t \bar{b}}
\,
m_{b}^2
\,
\Big[
g^{L}_{Z b \bar{b}}
\,
C_{1}
-
g^{R}_{Z b \bar{b}}
\,
\big(
C_{0}
+
C_{1}
+
C_{2}
\big)
\Big]
(Z,H^\pm,W;
b,b,t)
\nonumber \\
&&
-
m_{t}^2
g^{R}_{H^- t \bar{b}}
\Big(
g^{R}_{Z t \bar{t}}
\,
C_{1}
+
g^{L}_{Z t \bar{t}}
\,
C_{2}
\Big)
(W,Z,H^\pm;
b,t,t)
\nonumber\\
&&
\hspace{4cm}
-
m_{t}^2
g^{R}_{H^- t \bar{b}}
g^{L}_{Z b \bar{b}}
\,
C_{2}
(Z,H^\pm,W;
b,b,t)
\Big]
\Bigg\},
\nonumber
\end{eqnarray}
where the color index $N^C_Q$
for quarks such as
top quark $t$ and
bottom quark $b$ exchanging
in loop has a value of 3.
These related general couplings
for these vector
boson - fermion vertices
are expressed with
$g^{L}_{Z f \bar{f}}
= e/(s_{W} c_{W})
\times (I^3_f - s_{W}^2 Q_f),
\,\,\,
g^{R}_{Z f \bar{f}}
= e/(s_{W} c_{W})
\times (- s_{W}^2 Q_f)$
and
$g^{R}_{W^\pm f \bar{f'}}
= 0, \, g^{L}_{W^\pm f \bar{f'}}
\equiv g_{W^\pm f \bar{f'}}
= e/(\sqrt{2} s_W)$.
%
%
%
\subsection*{
\underline{Form factors
$\mathcal{T}^{F}_{i, \text{Self}}$}:
}
\begin{center}
\begin{figure}[H]
\centering
\includegraphics[width=10cm, height=3cm]
{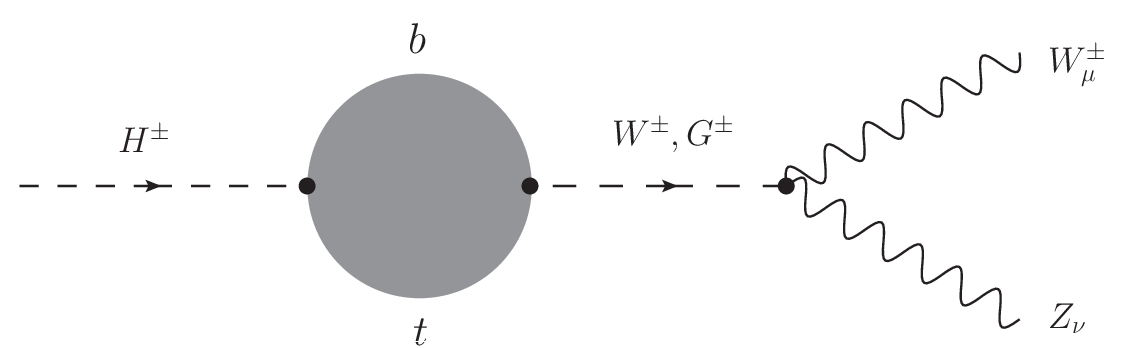}
\caption{
Self-energy Feynman diagram contributions
at the external leg $H^\pm$ with fermion loops.
}
\label{Figs:SelfF}
\end{figure}
\end{center}

The form factors
$\mathcal{T}^{F}_{i, \text{Self}}$
for $i = 1, 2, 3$,
which is arisen from
a typical topology in
Fig.~\ref{Figs:SelfF},
are presented as follows:
\begin{eqnarray}
\dfrac{
\mathcal{T}^{F}_{1, \text{Self}}
}{
g_{Z W^\pm W^\mp}
\cdot
g_{W^+ \bar{t} \, b}
}
&=&
\dfrac{
2 N^C_Q
}{
16\pi^2
\cdot
m_{W}^2
\big(
m_{H^\pm}^2 
- m_{W}^2\big
)
}
\Bigg\{
s_{W}^2
m_{Z}^2
\big(
m_{t}^2
\,
g^{R}_{H^- t \bar{b}}
-
m_{b}^2
\,
g^{L}_{H^- t \bar{b}}
\big)
A_{0}
(b)
\nonumber \\
&&
-
g^{L}_{H^- t \bar{b}}
\,
m_{b}^2
\,
\Big[
m_{W}^2
\big(
m_{Z}^2
-
m_{W}^2
\big)
-
s_{W}^2
m_{H^\pm}^2
m_{Z}^2
\Big]
\big(
B_{0}
+
B_{1}
\big)
(H^\pm;
b,t)
\nonumber \\
&&
+
s_{W}^2
m_{Z}^2
m_{t}^2
g^{R}_{H^- t \bar{b}}
\big(
m_{t}^2
-m_{b}^2
-m_{H^\pm}^2
\big)
B_{0}
(H^\pm; b,t)
\\
&&
+
m_{t}^2
g^{R}_{H^- t \bar{b}}
\Big[
m_{W}^2
\big(
m_{W}^2
-
m_{Z}^2
\big)
-
s_{W}^2
m_{H^\pm}^2
m_{Z}^2
\Big]
B_{1}
(H^\pm; b,t)
\Bigg\},
\nonumber \\
\mathcal{T}^{F}_{2, \text{Self}}
&=&
\mathcal{T}^{F}_{3, \text{Self}}
=
0.
\end{eqnarray}
The general couplings involving the Goldstone bosons
$G^\pm$ and quarks $f, f'$ exchanging in the loop
are given by $\lambda_{G^- f \bar{f'}}
= -i \Big( m_{f'} \, g^{L}_{G^- f \bar{f'}} P_L
+ m_f \, g^{R}_{G^- f \bar{f'}} P_R \Big)$,
and $\lambda_{G^+ f' \bar{f}}
= -i \Big( m_f \, g^{L}_{G^+ f' \bar{f}} P_L
+ m_{f'} \, g^{R}_{G^+ f' \bar{f}} P_R \Big)$.
Furthermore, the left- and right-handed
couplings satisfy the relations as
$g^{L}_{G^+ f' \bar{f}} = g^{R}_{G^- f \bar{f'}}
= -\frac{1}{m_W} \, g_{W^\pm f \bar{f'}}$ and
$g^{R}_{G^+ f' \bar{f}} = g^{L}_{G^- f \bar{f'}}
= +\frac{1}{m_W} \, g_{W^\pm f \bar{f'}}$.
\subsection*{
\underline{Form factors
$\mathcal{T}^{F}_{i, \text{Tad}}$}:
}
\begin{center}
\begin{figure}[H]
\centering
\includegraphics[width=14cm, height=5cm]
{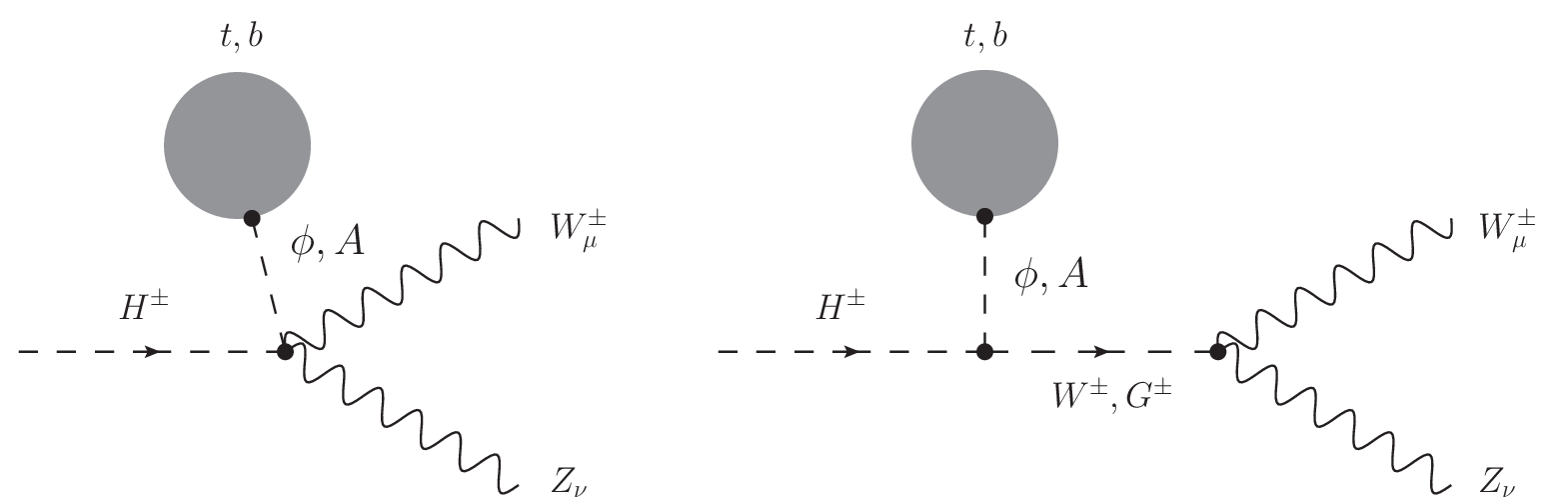}
\caption{
Tadpole Feynman diagram contributions
with poles $\phi \equiv h, H, A$ for
fermion loops.}
\label{Figs:TadpoleF}
\end{figure}
\end{center}

Regarding to Fig.~\ref{Figs:TadpoleF},
form factor
$\mathcal{T}^{F}_{i, \text{Tad}}$
for $i = 1, 2, 3$ is
expressed into one-loop
contributions by
pole $\phi \equiv h, H$
and pole $A$ as follows
\begin{eqnarray}
\mathcal{T}^{F}_{i, \text{Tad}}
&=&
\mathcal{T}^{F, A}_{i, \text{Tad}}
+
\sum
\limits_{\phi = h, H}
\mathcal{T}^{F, \phi}_{i, \text{Tad}}.
\end{eqnarray}
Where the related
general couplings for
these scalar Higgs $\phi$
and pseudo-scalar Higgs
$A$ with fermion $f$
vertices are introduced
with $g^{L}_{\phi f \bar{f}}
= g^{R}_{\phi f \bar{f}}
\equiv g_{\phi f \bar{f}}$
and $g^{R}_{A f \bar{f}}
= - g^{L}_{A f \bar{f}}
= g_{A f \bar{f}}$
as follows:
$
-i \,
m_f
\big(
g^{L}_{\phi f \bar{f}}
\,
P_L
+
g^{R}_{\phi f \bar{f}}
\,
P_R
\big)
=
-i \,
m_f
\,
g_{\phi f \bar{f}}$, and
$m_f
\big(
g^{L}_{A f \bar{f}}
\,
P_L
+
g^{R}_{A f \bar{f}}
\,
P_R
\big)
=
m_f
\,
g_{A f \bar{f}}
\,
\gamma_5$.
Therefore these all
form factors at pole
$A$ by an analytical
relation $g^{L}_{A f \bar{f}}
+ g^{R}_{A f \bar{f}} = 0$
and ones for $i = 2, 3$
at pole $\phi$ have
no contribution for
fermion tadpole diagrams,
\begin{eqnarray}
\label{tadB}
&&
\mathcal{T}^{F, A}_{1, \text{Tad}}
=
\mathcal{T}^{F, A / \phi}_{2, \text{Tad}}
=
\mathcal{T}^{F, A / \phi}_{3, \text{Tad}}
=
0
\end{eqnarray}
and thus, the remaining
form factor contributed
from only pole $\phi$
is reading as follows
\begin{eqnarray}
\mathcal{T}^{F, \phi}_{1, \text{Tad}}
&=&
\dfrac{
N^C_Q
}{
16\pi^2
\cdot
m_{\phi}^2
\big(m_{H^\pm}^2
- m_{W}^2 \big)
}
\times
\\
&&
\times
\Big[
\big(
m_{H^\pm}^2
-m_{W}^2
\big)
g_{\phi H^-W^+Z}
\nonumber \\
&&
\hspace{0.25cm}
+
\frac{s_W^2}{c_W^2}
\big(
m_{\phi}^2
-m_{H^\pm}^2
\big)
\;
g_{\phi H^- W^+}
\cdot 
g_{Z W^\pm W^\mp}
\nonumber\\
&&
\hspace{0.25cm}
+
\big(
m_{Z}^2
- m_{W}^2
\big)
g_{\phi H^- W^+}
\cdot 
g_{Z W^\pm W^\mp}
\Big]
\times
\nonumber
\\
&&
\times
\Big[
4 m_{b}^2
\,
g_{\phi b \bar{b}}
\,
A_{0}
(b)
+
4 m_{t}^2
\,
g_{\phi t \bar{t}}
\,
A_{0}
(t)
\Big].
\nonumber
\end{eqnarray}
\section*{
Appendix B: Analytical checks of
the $\xi$-gauge invariance}
In this Appendix, we perform the analytical
check for $\xi$-gauge invariance of these form factors.
It should be reminded that two gauge parameters $\xi_{V}$,
with $V = W, Z$, arise from the propagators of the
vector bosons $W^\pm$ and $Z$, the Nambu--Goldstone bosons
$G^\pm$ and $G^0$, and the corresponding ghost fields
$u_\pm$ and $u_Z$ in the general $R_\xi$ gauge.
Accordingly, we examine the one-loop form factors
$\mathcal{T}^{B}_i$ ($i = 1, 2, 3$), grouped by the
bosons exchanged in the loop. In particular,
$\mathcal{T}^{B}_2$ is taken as a representative example
for the analytical checks in our simplified demonstration.
The one-loop form factor $\mathcal{T}^{B}_2$ is as follows:
\begin{eqnarray}
\mathcal{T}^{B}_2
&=&
\mathcal{T}^{B}_{2, \text{Trig}}
+
\mathcal{T}^{B}_{2, \text{Self}}
+
\mathcal{T}^{B}_{2, \text{Tad}}.
\end{eqnarray}
In this case, the form factors
$\mathcal{T}^{B}_{2, \text{Self}}$ and
$\mathcal{T}^{B}_{2, \text{Tad}}$ do not contribute
in the boson-loop group, as explicitly shown in
the sections mentioned above [see Eqs.~(\ref{Eq:NiBSelf})
and (\ref{tadB})]. As a result, we focus only on
the form factor $\mathcal{T}^{B}_{2, \text{Trig}}$
in Eq.~(\ref{Eq:NiBTrig}), which can be expressed
in the concrete form as follows:
\begin{eqnarray}
\label{Eq:FFB2Trig}
\mathcal{T}^{B}_{2, \text{Trig}}
&=&
\dfrac{
1
}{
16\pi^2
\cdot
m_{H^\pm}^2
\cdot
\Lambda(H^\pm,W,Z)
}
\Big(
\mathcal{T}^{B, A}_{2, \text{Trig}}
	+
\mathcal{T}^{B, H^\pm}_{2, \text{Trig}}
	+
\mathcal{T}^{B, W^\pm}_{2, \text{Trig}}
	+
\mathcal{T}^{B, Z}_{2, \text{Trig}}
	+
\mathcal{T}^{B, W^\pm Z}_{2, \text{Trig}}
\Big).
\nonumber\\
\end{eqnarray}
There are five main contributions in which the
scalar Higgs $\phi$ appears together with charged
and neutral particles propagating in the loop.
These contributions are expressed in terms of the
scalar Passarino--Veltman functions $A_0$, $B_0$,
and $C_0$. They will be examined explicitly in
the following paragraphs for the analytical checks
of $\xi$-gauge invariance.

Next, we consider the remaining form factors  
$\mathcal{T}^{B, W^\pm}_{2, \text{Trig}},
\mathcal{T}^{B, Z}_{2, \text{Trig}},
\mathcal{T}^{B, W^\pm Z}_{2, \text{Trig}},$
and $\mathcal{T}^{B, A}_{2, \text{Trig}},$
which involve vector and Goldstone boson  
propagators in terms of  
the gauge parameters $\xi_{V}$  
for $V = W, Z$.  
First, we present the 
representative form for  
$\mathcal{T}^{B, W^\pm}_{2, \text{Trig}}$
with the cancellation checks of  
$\xi$-dependence as follows:
\begin{eqnarray}
\dfrac{
\mathcal{T}^{B, W^\pm}_{2,
\text{Trig}}
}
{
g_{Z W^\pm W^\mp}
}
&=&
\sum
\limits_{\phi = h, H}
\dfrac{
g_{\phi W^\pm W^\mp}
\cdot 
g_{\phi H^- W^+}
}{
2 c_{W}^2 
m_{W}^4
}
\Big\{
c_{W^\pm}^{0}
+
c_{W^\pm}^{1}
A_{0}(\phi)
+
c_{W^\pm}^{2}
A_{0}(W)
\nonumber \\
&&
+
c_{W^\pm}^{3}
B_{0}(W;\phi,W)
+
c_{W^\pm}^{4}
B_{0}(H^\pm;\phi,W)
+
c_{W^\pm}^{5}
B_{0}(Z;W,W)
\nonumber \\
&&
+
c_{W^\pm}^{6}
C_{0}(W,Z,H^\pm;\phi,W,W)
+
c_{W^\pm}^{7}
A_{0}(\xi_{W}\; W)
+
c_{W^\pm}^{8}
B_{0}(W;\phi,\xi_{W}\; W)
\nonumber \\
&&
+
c_{W^\pm}^{9}
B_{0}(H^\pm;\phi,\xi_{W} W)
+
c_{W^\pm}^{10}
B_{0}(Z;W,\xi_{W} W)
+
c_{W^\pm}^{11}
B_{0}(Z;\xi_{W}\; W,\xi_{W}\; W)
\nonumber \\
&&
+
c_{W^\pm}^{12}
C_{0}(W,Z,H^\pm;\phi,W,\xi_{W} W)
+
c_{W^\pm}^{13}
C_{0}(H^\pm,Z,W; \phi,W,\xi_{W} W)
\nonumber \\
&&
+
c_{W^\pm}^{14}
C_{0}(W,Z,H^\pm; \phi,\xi_{W} W,\xi_{W} W)
\Big\}.
\end{eqnarray}
Where the corresponding coefficients 
in the form factors are listed as 
follows:  
\begin{eqnarray}
c_{W^\pm}^{0}
&=&
c_{2W}
m_{H^\pm}^2 m_{W}^2  
(m_{\phi}^2-m_{H^\pm}^2-m_{W}^2) 
(m_{H^\pm}^2-m_{W}^2-m_{Z}^2) 
\cdot  
\Lambda(H^\pm,W,Z),
\nonumber \\
c_{W^\pm}^{1}
&=&
c_{2W}
m_{W}^2 
(m_{\phi}^2-m_{H^\pm}^2-m_{W}^2) 
(m_{H^\pm}^2-m_{W}^2+m_{Z}^2) 
\cdot  
\Lambda(H^\pm,W,Z),
\nonumber \\
c_{W^\pm}^{2}
&=&
-c_{W}^2 
(2 m_{W}^2-m_{Z}^2) 
(m_{\phi}^2-m_{H^\pm}^2-m_{W}^2) 
(m_{H^\pm}^2-m_{W}^2+m_{Z}^2) 
\cdot  
\Lambda(H^\pm,W,Z),
\nonumber \\
c_{W^\pm}^{3}
&=&
c_{W}^2
m_{H^\pm}^2 
m_{\phi}^4 
(2 m_{W}^2-m_{Z}^2) 
\Big[
	\big(
	m_{H^\pm}^2
	-
	m_{W}^2
	-
	m_{Z}^2
	\big)^2
	+
	8 m_{W}^2 m_{Z}^2
	\Big]
\nonumber \\
&&
-
c_{W}^2
m_{H^\pm}^2 
m_{\phi}^2 
(2 m_{W}^2-m_{Z}^2) 
\Big[
-2 m_{Z}^2 
\big(
m_{H^\pm}^4 
-5 m_{H^\pm}^2 m_{W}^2
-12 m_{W}^4
\big)
\nonumber \\
&&
\hspace{3cm}
+
m_{Z}^4 (m_{H^\pm}^2-3 m_{W}^2)
+
(m_{H^\pm}^2+3 m_{W}^2) 
(m_{H^\pm}^2-m_{W}^2)^2
\Big]
\nonumber\\
&&
+
2 c_{W}^2
m_{H^\pm}^2 
m_{W}^2 
\Big[
-m_{Z}^4 
\big(
m_{H^\pm}^4
+16 m_{H^\pm}^2 m_{W}^2
+7 m_{W}^4
\big)
\nonumber \\
&&
\hspace{2cm}
+
m_{Z}^2 (m_{H^\pm}^2+m_{W}^2) 
\big(
-m_{H^\pm}^4 
+12 m_{H^\pm}^2 m_{W}^2
+5 m_{W}^4
\big)
\nonumber \\
&&
\hspace{2cm}
+
2 m_{Z}^6 
(m_{H^\pm}^2+m_{W}^2)
-
2 m_{W}^2 (m_{H^\pm}^2-m_{W}^2)^2 
(m_{H^\pm}^2-3 m_{W}^2)
\Big]
\nonumber \\
&&
-
6 m_{H^\pm}^2 m_{W}^4 
\xi_{W}
\Big[
c_{W}^2 
(m_{W}^2-m_{Z}^2)
+
m_{W}^2 s_{W}^2
\Big]
(m_{\phi}^2-m_{H^\pm}^2) 
(m_{H^\pm}^2-m_{W}^2-m_{Z}^2),
\nonumber \\
c_{W^\pm}^{4}
&=&
c_{W}^2
m_{\phi}^4 
(m_{Z}^2-2 m_{W}^2)  
\Big[
2 m_{H^\pm}^6
+m_{H^\pm}^4 (3 m_{Z}^2
-5 m_{W}^2)
\\
&&
\hspace{4cm}
+2 m_{H^\pm}^2 
(m_{W}^2-m_{Z}^2) 
(2 m_{W}^2+3 m_{Z}^2)
-
(m_{W}^2-m_{Z}^2)^3
\Big]
\nonumber \\
&&
+
2 c_{W}^2 m_{\phi}^2 
(2 m_{W}^2-m_{Z}^2) 
\Big[
m_{H^\pm}^8+2 m_{H^\pm}^6 m_{Z}^2
-4 m_{H^\pm}^4 (m_{W}^2-m_{Z}^2)^2
\nonumber \\
&&
\hspace{1.5cm}
+
m_{H^\pm}^2 
\big(
4 m_{W}^6+3 m_{W}^4 m_{Z}^2
-8 m_{W}^2 m_{Z}^4+m_{Z}^6
\big)
-
m_{W}^2 
(m_{W}^2-m_{Z}^2)^3
\Big]
\nonumber \\
&&
+
c_{W}^2
m_{H^\pm}^8 
\big(
2 m_{W}^4+5 m_{W}^2 m_{Z}^2+m_{Z}^4
\big)
\nonumber\\
&&
-
2 c_{W}^2
m_{H^\pm}^6 
\big(
8 m_{W}^6+20 m_{W}^4 m_{Z}^2
-2 m_{W}^2 m_{Z}^4+m_{Z}^6
\big)
\nonumber \\
&&
+
c_{W}^2
m_{H^\pm}^4 
\big(
28 m_{W}^8-46 m_{W}^6 m_{Z}^2
+14 m_{W}^4 m_{Z}^4
+m_{W}^2 m_{Z}^6+m_{Z}^8
\big)
\nonumber \\
&&
+
c_{W}^2
m_{W}^2 
(m_{W}^2-m_{Z}^2) 
(2 m_{W}^2-m_{Z}^2) 
\Big[
m_{W}^2 (m_{W}^2-m_{Z}^2)^2
-2 m_{H^\pm}^2 
(4 m_{W}^4+m_{Z}^4)
\Big]
\nonumber \\
&&
-
2 m_{H^\pm}^2 m_{W}^2 
(m_{\phi}^2-m_{H^\pm}^2-m_{W}^2) 
\xi_{W}
\Big[
c_{W}^2 
(m_{W}^2-m_{Z}^2)
+
m_{W}^2 s_{W}^2
\Big]
\times
\nonumber \\
&&
\hspace{5cm}
\times 
\Big[
2 m_{H^\pm}^4-m_{H^\pm}^2
(m_{W}^2+m_{Z}^2)
-(m_{W}^2-m_{Z}^2)^2
\Big],
\nonumber \\
\dfrac{
c_{W^\pm}^{5}
}
{m_{H^\pm}^2 m_{W}^2 }
&=&
6 m_{\phi}^4 (2 m_{W}^2-m_{Z}^2) 
(m_{H^\pm}^2-m_{W}^2-m_{Z}^2)
\\
&&
+
m_{\phi}^2 
(2 m_{W}^2-m_{Z}^2) 
\times 
\nonumber\\
&&
\times
\Big[
m_{H^\pm}^2
\big(
8 m_{Z}^2
-10 m_{W}^2
-7 m_{H^\pm}^2
\big)
+
m_{W}^2
\big(
17 m_{W}^2
+8 m_{Z}^2
\big)
-m_{Z}^4
\Big]
\nonumber \\
&&
+
m_{H^\pm}^4 
\Big[
\big(
18 m_{W}^4-3 m_{W}^2 
m_{Z}^2+2 m_{Z}^4
\big)
-
m_{H^\pm}^2 
(2 m_{W}^2+m_{Z}^2)
\Big]
\nonumber \\
&&
+
m_{H^\pm}^2 
\Big[
m_{W}^2
\big(
18 m_{W}^4
+
17 m_{W}^2 m_{Z}^2
-
6 m_{Z}^4
\big)
-
m_{Z}^6
\Big]
\nonumber \\
&&
+
m_{W}^2 
\Big[
m_{W}^2
\big(
35 m_{W}^2 m_{Z}^2
-
34 m_{W}^4
-
16 m_{Z}^4
\big)
+
3 m_{Z}^6
\Big],
\nonumber \\
\dfrac{
c_{W^\pm}^{6}
}
{
2 c_{W}^2 m_{H^\pm}^2
}
&=&
3 m_{\phi}^6 m_{Z}^2 
(2 m_{W}^2-m_{Z}^2) 
(m_{H^\pm}^2-m_{W}^2-m_{Z}^2)
\\
&&
+
m_{\phi}^4 m_{Z}^2 
(2 m_{W}^2-m_{Z}^2) 
\times 
\nonumber\\
&&
\times 
\Big[
m_{H^\pm}^2
\big(
7 m_{Z}^2
-8 m_{W}^2
-5 m_{H^\pm}^2
\big)
+
m_{W}^2
\big(
13 m_{W}^2
+7 m_{Z}^2
\big)
-2 m_{Z}^4
\Big]
\nonumber \\
&&
+
m_{\phi}^2 
\Big[
m_{H^\pm}^6 
\big(
2 m_{W}^4+m_{W}^2 m_{Z}^2-2 m_{Z}^4
\big)
\nonumber\\
&&
+
m_{H^\pm}^4 
\big(
-6 m_{W}^6+23 m_{W}^4 m_{Z}^2
-13 m_{W}^2 m_{Z}^4+4 m_{Z}^6
\big)
\nonumber \\
&&
+ m_{H^\pm}^2 
\big(
6 m_{W}^8+15 m_{W}^6 m_{Z}^2
-12 m_{W}^4 m_{Z}^4
+8 m_{W}^2 m_{Z}^6-2 m_{Z}^8
\big)
\nonumber \\
&&
-
m_{W}^4 
\big(
2 m_{W}^6+39 m_{W}^4 m_{Z}^2
-27 m_{W}^2 m_{Z}^4+4 m_{Z}^6
\big)
\Big]
\nonumber \\
&&
+
m_{W}^2 
(m_{H^\pm}^2-m_{W}^2)
\times 
\nonumber\\
&&
\times 
\Big[
m_{H^\pm}^6 
(2 m_{W}^2+m_{Z}^2)
+
m_{H^\pm}^4 
\big(
-10 m_{W}^4-11 m_{W}^2 m_{Z}^2+m_{Z}^4
\big)
\nonumber \\
&&
+
m_{H^\pm}^2 
\big(
14 m_{W}^6-17 m_{W}^4 m_{Z}^2
+16 m_{W}^2 m_{Z}^4-2 m_{Z}^6
\big)
\nonumber \\
&&
-
m_{W}^2 
\big(
6 m_{W}^6+5 m_{W}^4 m_{Z}^2
-7 m_{W}^2 m_{Z}^4+2 m_{Z}^6
\big)
\Big],
\nonumber \\
c_{W^\pm}^{7}
&=&
-2 m_{H^\pm}^2 m_{W}^2 
\Big[
c_{W}^2 
(m_{W}^2-m_{Z}^2)
+
m_{W}^2 s_{W}^2
\Big]
\Lambda (H^\pm, W,Z)
=0
, 
\nonumber \\
\dfrac{c_{W^\pm}^{8}
}{
m_{H^\pm}^2 m_{W}^2 
}
&=&
\Big[
c_{W}^2 
(m_{W}^2-m_{Z}^2)
+
m_{W}^2 s_{W}^2
\Big] 
\times
\\
&&\times 
\Bigg\{
\Big[
m_{\phi}^2 
m_{H^\pm}^2
\big[
m_{H^\pm}^2
-
2 m_{W}^2 
(4-3 \xi_{W})
-
2 m_{Z}^2
\big]
+
2 m_{W}^2 m_{Z}^2
m_{\phi}^2 
(8-3 \xi_{W})
\nonumber \\
&&
\hspace{7cm}
+
m_{W}^4\; m_{\phi}^2 
(7-6 \xi_{W}) 
+
m_{Z}^4\; m_{\phi}^2 
\Big]
\nonumber\\
&&
+
m_{W}^2 m_{\phi}^2 
\Big[
-
m_{H^\pm}^4 (7 \xi_{W}+3)
+
2 m_{H^\pm}^2 
\big[
m_{W}^2 (\xi_{W}+9)
\nonumber \\
&&
+
m_{Z}^2 (4 \xi_{W}+3)
\big]
- 5 m_{W}^4 (3-\xi_{W})
+m_{W}^2 m_{Z}^2 (6-4 \xi_{W})
-m_{Z}^4 (\xi_{W}+3)
\Big]
\Bigg\}
=0
,
\nonumber \\
c_{W^\pm}^{9}
&=&
2
\Big[
c_{W}^2 
(m_{W}^2-m_{Z}^2)
+
m_{W}^2 s_{W}^2
\Big] (\xi_{W}-1) 
\times
\\
&&
\times 
 m_{H^\pm}^2 m_{W}^2 
(m_{\phi}^2-m_{H^\pm}^2)
\Big[
m_{H^\pm}^2 
\big(
2 m_{H^\pm}^2
-
m_{W}^2
-
m_{Z}^2
\big)
-
(m_{W}^2-m_{Z}^2)^2
\Big]
=0
, 
\nonumber \\
\dfrac{
c_{W^\pm}^{10}
}{m_{H^\pm}^2}
&=&
\Big[
c_{W}^2 
(m_{W}^2-m_{Z}^2)
+
m_{W}^2 s_{W}^2
\Big] 
\times
\\
&&\times 
\Bigg\{ 
4 m_{\phi}^2 
m_{Z}^2
\Big[
3 m_{\phi}^2  
(m_{Z}^2-m_{H^\pm}^2+m_{W}^2)
-
m_{Z}^2
(4 m_{H^\pm}^2+m_{W}^2)
\Big]
\nonumber \\
&&
+
2 m_{\phi}^2 m_{Z}^2 
\Big[
(m_{H^\pm}^2-m_{W}^2) 
(7 m_{H^\pm}^2+11 m_{W}^2)
+
m_{Z}^4
\Big]
-
2 m_{H^\pm}^6 m_{Z}^2
\nonumber \\
&&
+
m_{H^\pm}^2
m_{W}^2 
m_{Z}^2
\big(
2 m_{W}^2 
-
3 m_{H^\pm}^2 
\big)
+
m_{Z}^4 
\big(
2 m_{H^\pm}^2+m_{W}^2
\big) 
\big(
2 m_{H^\pm}^2-7 m_{W}^2
\big)
\nonumber \\
&&
-
m_{W}^2 
\big(
m_{H^\pm}^2-m_{W}^2
\big)^2 
\big(
4 m_{H^\pm}^2-5 m_{W}^2
\big)
+
m_{Z}^2
\big(
11 m_{W}^6 
-
2 m_{H^\pm}^2 m_{Z}^4
+
3 m_{W}^2 m_{Z}^4
\big)
\Bigg\}
\nonumber \\
&&-
\xi_{W}
m_{W}^2 
\Bigg\{
-2 m_{H^\pm}^2 
\Big[
3 m_{Z}^2 
(m_{\phi}^2+m_{Z}^2)
+
m_{W}^2 
(m_{W}^2
+
m_{Z}^2)
\Big]
\nonumber \\
&&
+
m_{W}^2 m_{Z}^2 
(6 m_{\phi}^2-5 m_{Z}^2)
+
6 m_{\phi}^2 m_{Z}^4
+
m_{H^\pm}^4 
(m_{W}^2+6 m_{Z}^2)
+
m_{W}^4
(m_{W}^2
+
4 m_{Z}^2)
\Bigg\}
\nonumber\\
&=&0,
\nonumber \\
c_{W^\pm}^{11}
&=&
\Big[
c_{W}^2 
(m_{W}^2-m_{Z}^2)
+
m_{W}^2 s_{W}^2
\Big] 
m_{H^\pm}^2 m_{Z}^2 
(m_{H^\pm}^2-m_{\phi}^2)
	\times
	\\
&&\times
	\Bigg[
	6
	(m_{\phi}^2 - m_{W}^2 \xi_{W})
	(-m_{H^\pm}^2+m_{W}^2+m_{Z}^2)
	-
	5 m_{W}^4
	\nonumber \\
	&&
\hspace{3cm}	
+   m_{H^\pm}^2
	(m_{H^\pm}^2
	+
	4 m_{W}^2
	-
	2 m_{Z}^2)
	+
	m_{Z}^2
	(4 m_{W}^2 
	+
	m_{Z}^2)
\Bigg]
=0
,
\nonumber \\
\dfrac{
c_{W^\pm}^{12}
}
{
2 m_{H^\pm}^2
}
&=&
\Big[
c_{W}^2 
(m_{W}^2-m_{Z}^2)
+
m_{W}^2 s_{W}^2
\Big] 
\Bigg\{
3 m_{\phi}^6 m_{Z}^2 
\big(
m_{W}^2+m_{Z}^2-m_{H^\pm}^2
\big)
\\
&&
	+
	m_{\phi}^4 
	\Big[
	m_{H^\pm}^4 
	(2 m_{W}^2+5 m_{Z}^2)
	\nonumber \\
	&&
\hspace{1cm}	
	+
	m_{H^\pm}^2 
	\big(
	-4 m_{W}^4+m_{W}^2 m_{Z}^2-7 m_{Z}^4
	\big)
	+
	2 (m_{W}^2-m_{Z}^2)^2 (m_{W}^2+m_{Z}^2)
	\Big]
\nonumber \\
&&
-
m_{\phi}^2 m_{H^\pm}^2 
\Big[
m_{H^\pm}^4 (3 m_{W}^2+2 m_{Z}^2)
\nonumber\\
&& 
\hspace{2cm}
+
m_{H^\pm}^2 
\big(-3 m_{W}^4 
+5 m_{W}^2 m_{Z}^2
-4 m_{Z}^4 \big)
+
2 m_{Z}^2 (m_{W}^2-m_{Z}^2)^2
\Big]
\nonumber\\
&&
	+
	m_{H^\pm}^4 m_{W}^2 
	\Big[
	m_{H^\pm}^4+m_{H^\pm}^2 
	(m_{W}^2+m_{Z}^2)
	-2 (m_{W}^2-m_{Z}^2)^2
	\Big]
\Bigg\}
\nonumber \\
&&
- (2 \xi_{W})
m_{W}^2 
	(m_{\phi}^2-m_{H^\pm}^2) 
	\times 
	\nonumber\\
	&& 
	\times 
\Bigg\{
	m_{\phi}^2 
	\Big[
	m_{H^\pm}^4
	-
	2 m_{H^\pm}^2 
	(m_{W}^2+m_{Z}^2)
	+
	m_{W}^2
	(m_{W}^2
	+
	4 m_{Z}^2)
	+
	m_{Z}^4
	\Big]
	\nonumber \\
	&&
\hspace{3cm}	
	+
	m_{W}^2 
	\Big[
	m_{H^\pm}^2 
	(m_{W}^2+m_{Z}^2
	-2 m_{H^\pm}^2)
	+
	(m_{W}^2-m_{Z}^2)^2
	\Big]
	\Bigg\}
=0,
\nonumber \\
\dfrac{
c_{W^\pm}^{13}
}
{
2 m_{H^\pm}^2
}
&=&
\Big[
c_{W}^2 
(m_{W}^2-m_{Z}^2)
+
m_{W}^2 s_{W}^2
\Big]
\Bigg\{
3 m_{\phi}^6 m_{Z}^2 
\big(
-m_{H^\pm}^2+m_{W}^2+m_{Z}^2
\big)
\\
&&
+
m_{\phi}^4 
\Big[
	m_{H^\pm}^4 
	(5 m_{Z}^2-2 m_{W}^2)
	+
	m_{H^\pm}^2 
	\big(
	4 m_{W}^4+6 m_{W}^2 m_{Z}^2-7 m_{Z}^4
	\big)
	\nonumber \\
	&&
	\hspace{6cm}
	+
	2 m_{Z}^6
	-
	2 m_{W}^6
	-
	3 m_{W}^2 m_{Z}^2
	(5 m_{W}^2 
	+
	m_{Z}^2)
	\Big]
	\nonumber \\
&&
	+
	m_{\phi}^2 
	\Big[
	m_{H^\pm}^4 
	\big(
	6 m_{W}^4-4 m_{W}^2 m_{Z}^2+4 m_{Z}^4
	-2 m_{H^\pm}^2 m_{Z}^2
	\big)
\nonumber\\
&&
\hspace{2cm}
	-
	m_{H^\pm}^2 
	\big(
	15 m_{W}^6
	-5 m_{W}^4 m_{Z}^2
	+m_{W}^2 m_{Z}^4
	+2 m_{Z}^6
	\big)
\nonumber \\
	&&
\hspace{5cm}
	+
	m_{W}^2 
	\big(
	9 m_{W}^6+16 m_{W}^4 m_{Z}^2
	-8 m_{W}^2 m_{Z}^4+m_{Z}^6
	\big)
	\Big]
	\nonumber \\
	&&
	-
	m_{W}^2 
	\Big[
	m_{H^\pm}^4 
	\big(
	5 m_{W}^4+3 m_{W}^2 m_{Z}^2+m_{Z}^4
	-2 m_{H^\pm}^2 m_{Z}^2
	\big)
	\nonumber \\
	&&
	\hspace{2cm}
	+
	m_{H^\pm}^2 
	\big(
	-16 m_{W}^6
	+5 m_{W}^4 m_{Z}^2
	-3 m_{W}^2 m_{Z}^4+m_{Z}^6
	\big)
	\nonumber \\
	&&
	\hspace{5cm}
	+
	m_{W}^2 
	\big(
	11 m_{W}^6-8 m_{W}^4 m_{Z}^2
	+4 m_{W}^2 m_{Z}^4-m_{Z}^6
	\big)
	\Big]
\nonumber \\
&&
+
\xi_{W}
m_{W}^2  
	\Bigg[
	2 m_{\phi}^4 
	\Big[
	m_{Z}^2 
	(m_{H^\pm}^2+m_{W}^2-2 m_{Z}^2)
	+
	(m_{H^\pm}^2-m_{W}^2)^2
	\Big]
	\nonumber \\
&&
-
m_{\phi}^2 
\Big[
m_{H^\pm}^4 
(m_{Z}^2-5 m_{W}^2+3 m_{H^\pm}^2)
\nonumber\\
&&
\hspace{2cm}
+
m_{H^\pm}^2 
\big( 
-5 m_{W}^
+14 m_{W}^2 m_{Z}^2
-5 m_{Z}^4
\big)
\nonumber \\
&&
\hspace{4cm}
+
(m_{W}^2-m_{Z}^2) 
(7 m_{W}^2-m_{Z}^2) 
(m_{W}^2+m_{Z}^2)
\Big]
\nonumber \\
&&
-
m_{Z}^2 
(m_{H^\pm}^2-3 m_{W}^2) 
(m_{H^\pm}^4-3 m_{W}^4)
\nonumber \\
&&
\hspace{2cm}
+
m_{H^\pm}^4
(m_{H^\pm}^4
-
8 m_{W}^4)
+
m_{Z}^6 
(m_{H^\pm}^2-m_{W}^2)
\nonumber\\
&&
\hspace{4cm}
-
m_{Z}^4 
(m_{H^\pm}^2-3 m_{W}^2) 
(m_{H^\pm}^2+m_{W}^2)
+
7 m_{W}^8
\Bigg\}=0, 
\nonumber\\
\dfrac{
c_{W^\pm}^{14}
}
{
2 m_{H^\pm}^2
}
&=&
\Big[
c_{W}^2 
(m_{W}^2-m_{Z}^2)
+
m_{W}^2 s_{W}^2
\Big] 
\times
\\
&&\times 
\Bigg\{
m_{Z}^2 (m_{\phi}^2-m_{H^\pm}^2) 
\Bigg[
3 m_{\phi}^4 
\big( m_{H^\pm}^2
-m_{W}^2-m_{Z}^2
\big)
\nonumber\\
&&
\hspace{1.5cm}
-
2 m_{\phi}^2 
\Big[
m_{H^\pm}^2 
(m_{H^\pm}^2+m_{W}^2-2 m_{Z}^2)
-
2 m_{W}^4
+
m_{Z}^2
(m_{W}^2 + m_{Z}^2)
\Big]
\nonumber \\
&&
\hspace{4cm}
-
m_{W}^2 
\Big[
m_{H^\pm}^2 
(m_{W}^2+m_{Z}^2-2 m_{H^\pm}^2)
+
(m_{W}^2-m_{Z}^2)^2
\Big]
\Bigg]
\nonumber \\
&&
-
(\xi_{W})
m_{W}^2 
(m_{\phi}^2-m_{H^\pm}^2) 
\times 
\nonumber\\
&&
\times 
\Bigg[
m_{Z}^4 
\big(-6 m_{\phi}^2
+m_{H^\pm}^2
-3 m_{W}^2
-m_{Z}^2
\big)
\nonumber \\
&&
+
m_{Z}^2 
(m_{H^\pm}^2-m_{W}^2) 
(6 m_{\phi}^2+m_{H^\pm}^2-3 m_{W}^2)
-
(m_{H^\pm}^2-m_{W}^2)^3
\Bigg]
\Bigg\}
=0.
\nonumber
\end{eqnarray}
It is noted that the kinematical function is
defined as
$\Lambda(H^\pm, W, Z) = (m_{H^\pm}^2
- m_W^2 - m_Z^2)^2 - 4 m_W^2 m_Z^2$.
Because $c_W^2 (m_W^2 - m_Z^2) + m_W^2 s_W^2 = 0$,
the coefficients $c_{W^\pm}^{7} = c_{W^\pm}^{8}
= \cdots = c_{W^\pm}^{14} = 0$.
While $c_{W^\pm}^{3}$ and $c_{W^\pm}^{4}$ are independent of $\xi$,
the remaining coefficients $c_{W^\pm}^{1,2}$ and $c_{W^\pm}^{5,6}$
also do not depend on $\xi$. Consequently, the form factor
$\mathcal{T}^{B}_{2, \text{Trig}}$ becomes $\xi$-independent.
Other form factors are confirmed using the same procedure,
demonstrating that they are also $\xi$-independent.
\section*{
Appendix C: Feynman diagrams for
$H^{\pm}\rightarrow W^{\pm}Z$ in
the general $R_{\xi}$}
A complete set of one-loop 
Feynman diagrams relevant 
to the decay process 
$H^{\pm}\rightarrow W^{\pm}Z$ 
in the general $R_{\xi}$ gauge 
is provided in the Appendix.

\begin{center}
\begin{figure}[H]
\centering
\includegraphics[width=10cm, height=4cm]
{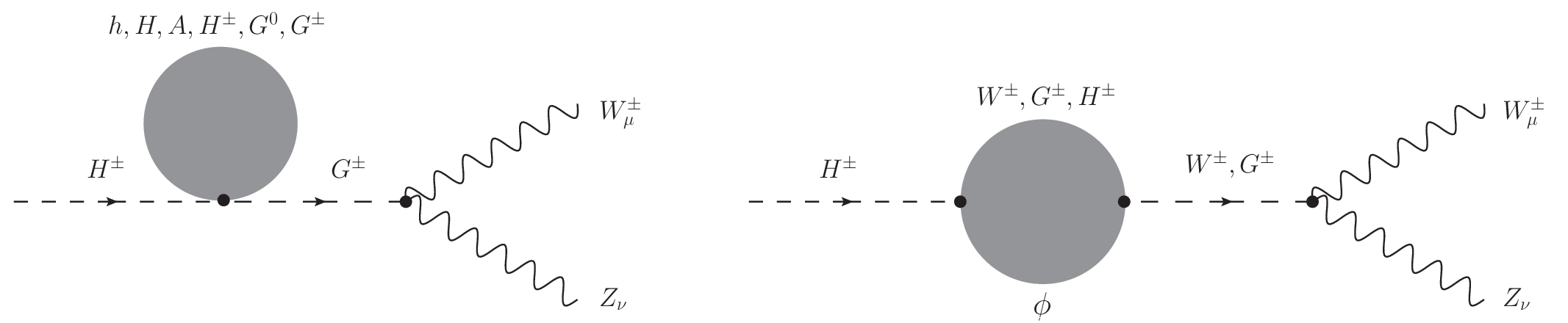}
\caption{
Self-energy Feynman diagram contributions to the external leg $H^\pm$.}
\label{Figs:SelfB}
\end{figure}
\end{center}

\end{document}